# Time and position distributions in large volume spherical scintillation detectors

Paolo Lombardi and Gioacchino Ranucci*
*Istituto Nazionale di Fisica Nucleare*
*Via Celoria 16, 20133 Milano, Italy*
*e-mail: paolo.lombardi@mi.infn.it; gioacchino.ranucci@mi.infn.it*

ABSTRACT

Large spherical scintillation detectors are playing an increasingly important role in experimental neutrino physics studies. From the instrumental point of view the primary signal response of these set-ups is constituted by the time and amplitude of the anode pulses delivered by each individual phototube following a particle interaction in the scintillator. In this work, under some approximate assumptions, we derive a number of analytical formulas able to give a fairly accurate description of the most important timing features of these detectors, intended to complement the more complete Monte Carlo studies normally used for a full modelling approach. The paper is completed with a mathematical description of the event position distributions which can be inferred, through some inference algorithm, starting from the primary time measures of the photomultiplier tubes.

*Corresponding author. Tel: +39-02-50317362. E-mail address: gioacchino.ranucci@mi.infn.it.

## 1. Introduction

Contemporary large liquid scintillator detectors represent the latest evolution of a mature technology which has been, since its introduction, a key player in the particle detection arena. The rather easy possibility to accumulate large detection mass at not prohibitive costs makes the choice of massive scintillation detectors particularly attractive for the field of neutrino physics, where indeed a variety of scintillator set-ups for supernova neutrino detection [1] and for reactor neutrino oscillations [2][3] [4] [5] have been designed, installed and operated over the past two decades. Furthermore, in the solar neutrino field the massive spherical detectors Borexino [6] (based upon the fundamental pioneering experience of its pilot prototype Counting Test Facility [7]) and KamLAND (solar phase) [8] are actively pushing toward an imminent start of data taking in the background challenging sub-MeV region, while ambitious plans are being shaped to ensure a future to the SNO apparatus as a multipurpose scintillator neutrino observatory (SNO+) [9].

In the early days of the development of the scintillation technique a big deal of studies were devoted to the theoretical and experimental investigation of the timing features of scintillation counters (i.e. assemblies of scintillators and phototubes), reaching a fairly accurate description of the timing properties of counters of modest size (see for example [10][11] [12]).

Despite the complexity associated with the size, the signal characterization in large set-ups in principle can follow approaches similar to those adopted in small scintillation equipments, since also large detectors comprise essentially, in this respect, the same two major components, i.e. the scintillator in which the interactions occur and through which the light propagates, and the phototubes dedicated to the acquisition of the scintillation photons.

From a pure instrumental point of view, the detector signal response to the interactions of interest corresponds to the ensemble of photoelectron signals delivered by each individual phototube: specifically, the two quantities of relevance characterizing the output pulses are the number of photoelectrons by which they are formed and the timing evaluated with respect to a common trigger pulse. From these primary measures then the physical quantities of interest of each interaction in the detector are computed, essentially the energy of the events and their spatial location.

The full understanding of the detector response implies normally an integrated twofold approach: a thorough Monte Carlo modelling, as accurate as possible, of the light production and transport phenomena from the interaction site to the detecting photomultipliers, complemented by an "in-situ" source calibration of the detector. In the particular case of special interest of spherical geometry, a preliminary, basic understanding of the overall detector response can be, however, gained through some simple geometric formulas which, although cannot replace the need for a complete Monte Carlo modelling, can help to get an immediate realistic picture of the performance of the whole device, especially for what concerns the time distributions of the phototube pulses. Purpose of this work is to show in details the derivation of these formulas, which practically represent a sort of "zero order" description of the performances of a large spherical scintillation counter.

The paper is organized as follows : since it has been mentioned that the phototube timing in a real set-up is related to a common trigger, as introduction to the whole problematic in paragraph 2 the well known issue of the random trigger rate from a coincidence arrangement of a plurality of PMT's is reviewed; then in paragraph 3 some general considerations on the timing in large scintillation counters are given, defining also the limit of validity of the derivations presented in the subsequent paragraphs. In paragraph 4 it is considered the timing response of a single PMT observing a spherical detector characterized by a uniform distribution of events, while in paragraph 5 the case of a uniform surface distribution of events is considered. In paragraph 6 it is illustrated in detail the special case in which the detector is characterized by an index of refraction discontinuity, as for example when the scintillation vessel is surrounded by a shielding liquid of different nature. In paragraph 7 it is addressed the complementary issue of the photoelectron time sequence detected by the phototube lattice as response to an event occurring in a generic location inside the detector, being the calculation done in both cases of single refraction index and two refraction indexes. In paragraph 8 the description of the timing features is completed with a digression concerning the expected fluctuation of the trigger pulse derived from the coincidence criterion of $N_F$ PMT's firing in a predefined window.

In the second part of the paper the question is faced of the estimated radial distributions of the event locations, as derived from the primary time distributions. Actually, this problem is addressed without considering in details the procedure to infer the spatial location estimate from the time information (for this purpose see for example [13]), but it is assumed that the estimation process results in Gaussian resolution functions affecting equally each individual spatial coordinate ($x$, $y$, $z$), and it is then illustrated how such estimates combine to produce the estimated radial event distributions in various cases of interest. Preliminarily, in paragraph 9 it is demonstrated what are, individually, the global $x$, $y$ and $z$ expected distributions in case of a class of events uniformly distributed in the vessel. Then in paragraph 10 the estimated radial profile of point like events is derived, while in paragraph 11 the same type of distribution is obtained for uniformly distributed events, and in paragraph 12 for external events. Finally, in paragraph 13 there are the conclusions.

## 2. Random coincidence rate

The trigger condition in a large self-triggering scintillation detector viewed by $N_T$ PMT's is normally set by requiring a certain number $N_F$ of concurrently firing phototubes within a predefined time window of length $T$. Such a length is defined to be comparable with the photon time of flight across the whole volume of the detector, while $N_F$ is fixed through a trade-off between the minimum desirable energy to be observed and the rate of the random triggers due to the fake coincidences originated by the dark noise pulses emitted by the tubes. The resulting random trigger rate may pose a problem when the phototubes work in the single photoelectron regime, since in that case the noise pulses and the signal pulses are indistinguishable. In order to fix properly $N_F$ it is needed to know the formula which governs the rate of false occurring triggers.

Since a trigger occurs when $N_F$ out of the total $N_T$ devices fire, we have as many trigger configurations as the number of the groups of $N_T$ PMT's grouped in classes of dimension $N_F$, i.e. $\begin{pmatrix} N_T \\ N_F \end{pmatrix}$. Assuming for simplicity that all the PMT's feature the same dark noise rate $r$, then for each configuration we have, given a PMT firing as first among the $N_F$ of that configuration, that the probability of the remaining *(NF-1)* PMT's to fire within a time window $T$ from the first is $r^{N_F-1}T^{N_F-1}$, from which by multiplying by $r$ we get the rate, i.e. $r^{N_F}T^{N_F-1}$. Actually, this is the rate due to the considered configuration, but with one specific PMT firing as first; the overall rate is thus obtained considering that any of the $N_F$ tubes can be the first, which implies that the previous rate is multiplied by $N_F$, i.e. $N_F r^{N_F}T^{N_F-1}$. Finally, taking into account all the possible PMT's configurations, one obtains as overall random coincidence rate $R$

$$R = \begin{pmatrix} N_T \\ N_F \end{pmatrix} N_F r^{N_F} T^{N_F-1} \qquad (1).$$

Essentially the same formula can be found, for example, in [14].

For a given number of total PMT's $N_T$, Eq. (1) is a sharply varying function both of the number $N_F$ of phototubes in trigger and of the rate $r$. This is clearly demonstrated in Fig. 1, where we plot the relation (1) for a detector characterized by $N_T$ equal to 2200 and by a coincidence time $T$ equal to 50 ns (for a realistic example, these numbers adhere to the features of the solar neutrino Borexino detector [6]). The five curves in the figure, which correspond to five different values of the PMT's dark rate (chosen in the more likely range of the Borexino PMT's), i.e. 0.5, 1, 2, 3 and 4 kHz, give the random trigger rate expressed in events per day for an increasing number of PMT's in trigger from 1 to 20. Initially the rate is extremely high, since in the naïve condition of only 1 PMT in trigger the random rate is obviously equal to the number of PMT's multiplied by the individual dark rate; the random rate becomes negligible even for the highest rate of 4 kHz, however, already with only 14-15 PMT's in trigger.

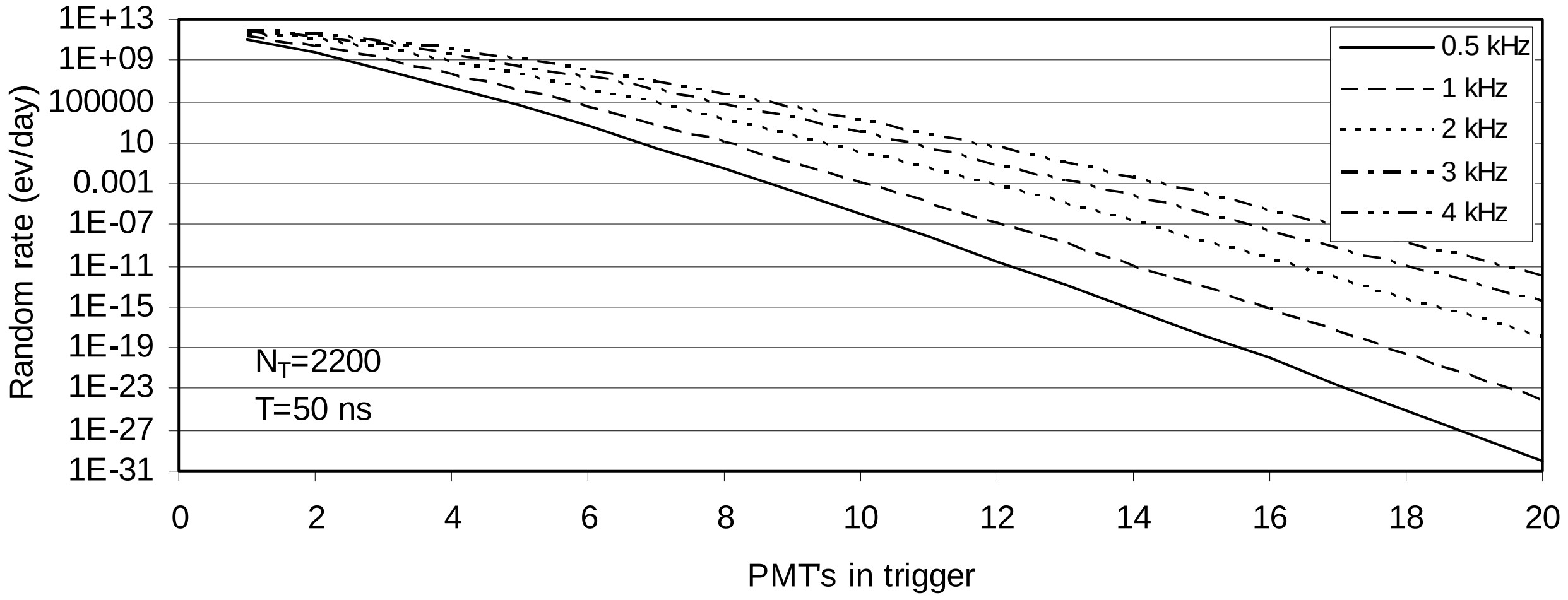

*Fig. 1 - Random coincidence trigger rate as function of the number of PMT's in coincidence and for different dark rates of the phototubes*

This result implies that the threshold imposed by the requirement to keep to a minimum the random trigger rate does not represent, in practical case, a real limitation on the low energy response of a scintillation detector.

**3. General timing considerations in large scintillation counters**

As mentioned in the introduction, when referring to the timing features in scintillation counters one implicitly means the characteristics of the time distributions of the photoelectrons revealed by the observing phototubes as response to an interaction occurring inside the scintillator. These features depend upon a multiplicity of factors: the intrinsic scintillation decay time of the scintillator [10], the time of flight (TOF) of the photons through the detector prior to arrive to the phototubes, which takes into account the geometrical extension of the detector itself, as well as the attenuation, absorption re-emission and scattering effects along the photon path, and finally the timing in the photomultipliers along the conversion process of the photons into photoelectrons and the subsequent multiplication throughout the dynode chain [15].

It should be reminded that the absorption re-emission effect stems from the overlap of the emission and absorption spectra of the scintillation medium, so that a photon can undergo self-absorption followed by a subsequent re-emission [16]. The process is quite complex, and its main effect is to enhance the tails of the timing distributions. In previous works [17] and [18] it was shown a mathematical procedure able to deal with the absorption re-emission effect and to quantify its impact on the overall timing properties. On the contrary, in this work the focus is on the determination of the TOF geometrical factors in different hypotheses of intrinsic event distributions, but under the simplifying assumption of neglecting the absorption re-emission effect (as well as the scattering), and including at most the simple attenuation effect.

Since the overall photoelectron time at the output of each phototube is the sum of the factors listed above, its global probability density function (PDF) is the convolution [19] of the individual distributions pertaining to each factor. For the purpose of the examples in this work, the scintillation decay profile is assumed equal to that of the scintillation mixture (Pseudocumene+PPO at 1.5g/l) adopted for the Borexino solar neutrino experiment, while the PMT time response is modeled as a Gaussian distribution with sigma equal to 1 ns. In the following paragraphs we will always show the TOF factor resulting from the calculations related to the various cases under consideration, and, only when useful to illustrate better the obtained results, also the convolution of the derived TOF factor with the scintillation curve and the PMT response. It will be also clear, along the calculations, that the simple attenuation effect, when considered, will imply essentially a modification of the TOF term.

**4. Time distribution detected by a phototube for a uniform volume distribution of events**

Let's consider (Fig. 2) a spherical volume of radius $R$ of liquid scintillator with events uniformly distributed in it. We want to evaluate the time distribution that such events induce on a PMT observing the volume, neglecting the absorption re-emission and scattering effects. If we consider a polar coordinate system centered on the PMT, given a point characterized by coordinates $l,\ \theta$ ($\theta$ is the angular coordinate orthogonal to the plane of the figure and this is why it is not reported in the figure itself) and $\varphi$, we have that the probability that a scintillation event occurs in the infinitesimal volume around such a point (shaded area in the figure) is

$$\frac{1}{\frac{4}{3}\pi R^3} l^2 \sin\varphi d\varphi d\theta dl \qquad (2)$$

and the probability that a photon emitted from this point hits the phototube is

$$\frac{dS}{4\pi l^2}\cos(\pi-\varphi) \qquad (3).$$

Hence the joint probability that a photon is generated at the location $l,\ \theta$ and $\varphi$ and is detected by the phototube under consideration is

$$\frac{1}{\frac{4}{3}\pi R^3} l^2 \sin\varphi d\varphi d\theta dl \frac{dS}{4\pi l^2}\cos(\pi-\varphi) \qquad (4).$$

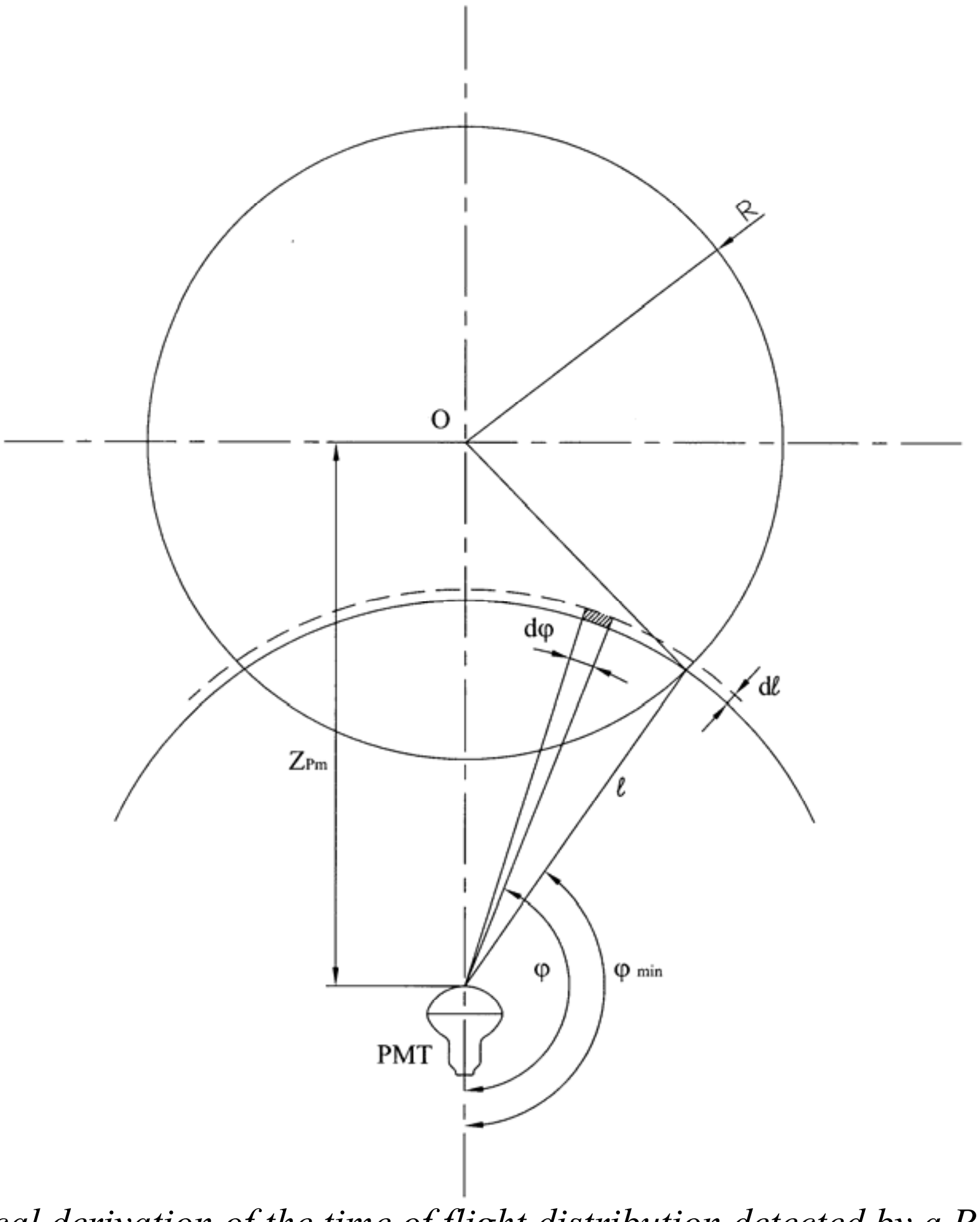

*Fig. 2 – Geometrical derivation of the time of flight distribution detected by a PMT observing a spherical detector characterized by a uniform internal distribution of scintillation events*

The integration of relation (4) over $\theta$ and $\varphi$ gives thus the PDF of the path traveled by the photons arriving at the phototube. Hence

$$p(l) = \frac{1}{\frac{4}{3}\pi R^3} \int_{\varphi=\varphi_{\min}}^{\pi} \int_{\theta=0}^{2\pi} l^2 \sin\varphi d\varphi d\theta dl \frac{dS}{4\pi l^2}\cos(\pi-\varphi) d\theta d\varphi \qquad (5)$$

where $\varphi_{min}$ is the minimum $\varphi$ that, for each $l$, corresponds to a real path from the scintillator sphere to the PMT (see Fig. 2).

Eq. (5) becomes

$$p(l) = \frac{1}{\frac{4}{3}\pi R^3} \int_{\varphi=\varphi_{\min}}^{\pi} \int_{\theta=0}^{2\pi} l^2 \sin\varphi d\varphi d\theta dl \frac{dS}{4\pi l^2}\cos(\pi-\varphi) d\theta d\varphi \qquad (6)$$

or

$$p(l) = \frac{2\pi}{\frac{4}{3}\pi R^3} \int_{\varphi=\varphi_{\min}}^{\pi} \frac{dS}{4\pi}(-\sin\varphi\cos\varphi) d\varphi \qquad (7)$$

$$p(l) = \frac{1}{\frac{8}{3}\pi R^3} \int_{\varphi=\varphi_{\min}}^{\pi} -\frac{1}{2}\sin 2\varphi d\varphi dS \qquad (8).$$

Eq. (8) after some manipulations becomes finally

$$p(l) = \frac{3}{16\pi R^3}\left(1 - \frac{\left(R^2 - l^2 - Z_{pm}^2\right)^2}{4 l^2 Z_{pm}^2}\right) dS \qquad (9).$$

In order to obtain the relation (9), it is taken into account that, from the triangle formed by $R$, $l$, and $Z_{pm}$ (see again Fig. 2), the angle $\varphi_{min}$ is given by

$$\cos\varphi_{\min} = \frac{R^2 - l^2 - Z_{pm}^2}{2 l Z_{pm}} \qquad (10).$$

One may wonder which is the meaning of the surface element $dS$ which appears in (9). Specifically, it means that the integral of (9) over $l$ gives the probability to detect a photon over the unit surface in correspondence to the phototube, so that the further integral over the whole detecting surface (i.e. the sphere of radius $Z_{pm}$ covered by the observing phototubes) gives 1. Since the detection probability for unit surface, for obvious symmetry reasons, is uniform over the entire detecting sphere, such an integral is equivalent to multiply by the area of the sphere of radius $Z_{pm}$ (we do not consider here the obvious fact that in a practical arrangement the detecting sphere is only partially covered by the lattice of PMT's).

The above calculation neglects the attenuation. For a more realistic result also the self-attenuation effect of the scintillator should be considered. Assuming, as it is in many actual set-ups, that the same attenuation process occurs inside the vessel and in the path from the vessel to the phototube, it is enough to consider a multiplicative exponential factor so that eq. (9) becomes

$$p(l) = \frac{3}{16\pi R^3}\left(1 - \frac{\left(R^2 - l^2 - Z_{pm}^2\right)^2}{4l^2 Z_{pm}^2}\right) e^{-\frac{l}{l_{att}}} dS \qquad (11)$$

where $l_{att}$ is the attenuation length.

Eq. (9) (and similarly eq. (11)) can be expressed in term of the variable time of flight $t_f$, i.e. remembering that

$$l = \frac{c}{n} t_f$$

we have

$$p(t_f) = \frac{3}{16\pi R^3}\left(1 - \frac{\left(R^2 - \left(\frac{c}{n}t_f\right)^2 - Z_{pm}^2\right)^2}{4\left(\frac{c}{n}t_f\right)^2 Z_{pm}^2}\right) \frac{c}{n} e^{-\frac{\frac{c}{n}t_f}{l_{att}}} dS \qquad (12)$$

where $c/n$ is the Hessian of the transformation.

The above relations are valid for $Z_{pm} - R < l < Z_{pm} + R$ (i.e. for $\frac{n}{c}\left(Z_{pm} - R\right) < t_f < \frac{n}{c}\left(Z_{pm} + R\right)$).

The time distribution for unit surface (eq. (12) omitting the factor $dS$) is plotted in Fig. 3 for $R=4.25$ and $Z_{pm}=6.52$; these geometrical parameters correspond to those of the solar neutrino experiment Borexino. The two plots correspond respectively to either ignoring or including the attenuation effect, with an attenuation length taken for the purpose of the example equal to 7 m. Since in Borexino the inner containment vessel will be immersed in the same solvent used as base of the scintillator (Pseudocumene) the latter calculation is that more realistic.

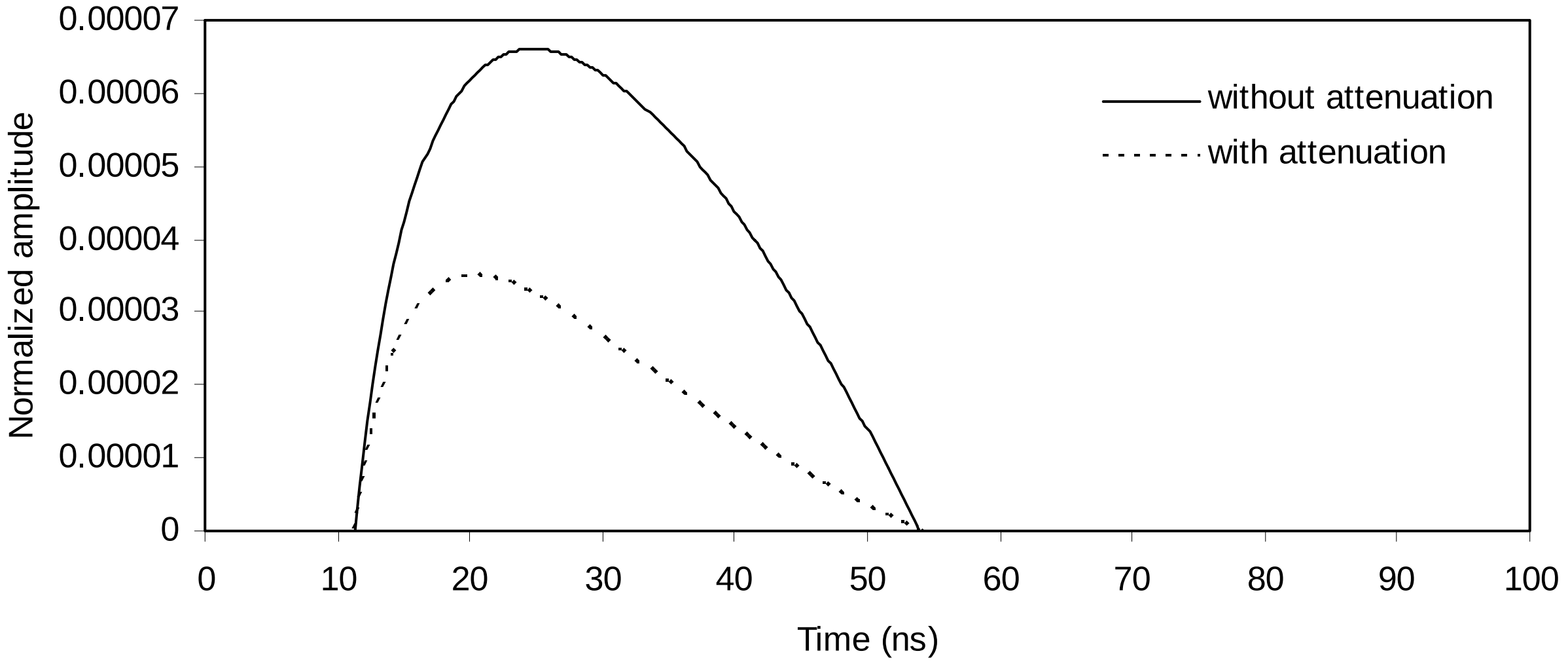

*Fig. 3 - Time of flight distribution induced on an observing phototube by a uniform distribution of events in the scintillation vessel*

**5. Time distribution detected by a phototube for a uniform surface distribution of events**

The evaluation of the time of flight distribution induced by a uniform surface distribution of events proceeds similarly to the procedure illustrated in the previous paragraph. With reference to Fig. 4 one can note that the events characterized by a same path to arrive to the PMT are those comprised in the elementary surface subtended by the angle $d\theta$. Taking into account the circular symmetry of the problem, the area of this surface can be written as $2\pi R^2 \sin\theta d\theta$. Hence the joint probability that a photon is emitted from the surface subtended by $d\theta$ and that is detected by the phototube (corresponding to an elementary detection area $dS$) is

$$p(l)dldS = \frac{1}{4\pi R^2} 2\pi R^2 \sin\theta d\theta \frac{dS}{4\pi l^2} \cos\varphi \quad (13).$$

The angle $\theta$ can be expressed as function of $l$ exploiting the cosine theorem for the triangle OP(PMT). i.e.

$$\cos\theta = \frac{Z_{pm}^2 + R^2 - l^2}{2RZ_{pm}} \quad (14).$$

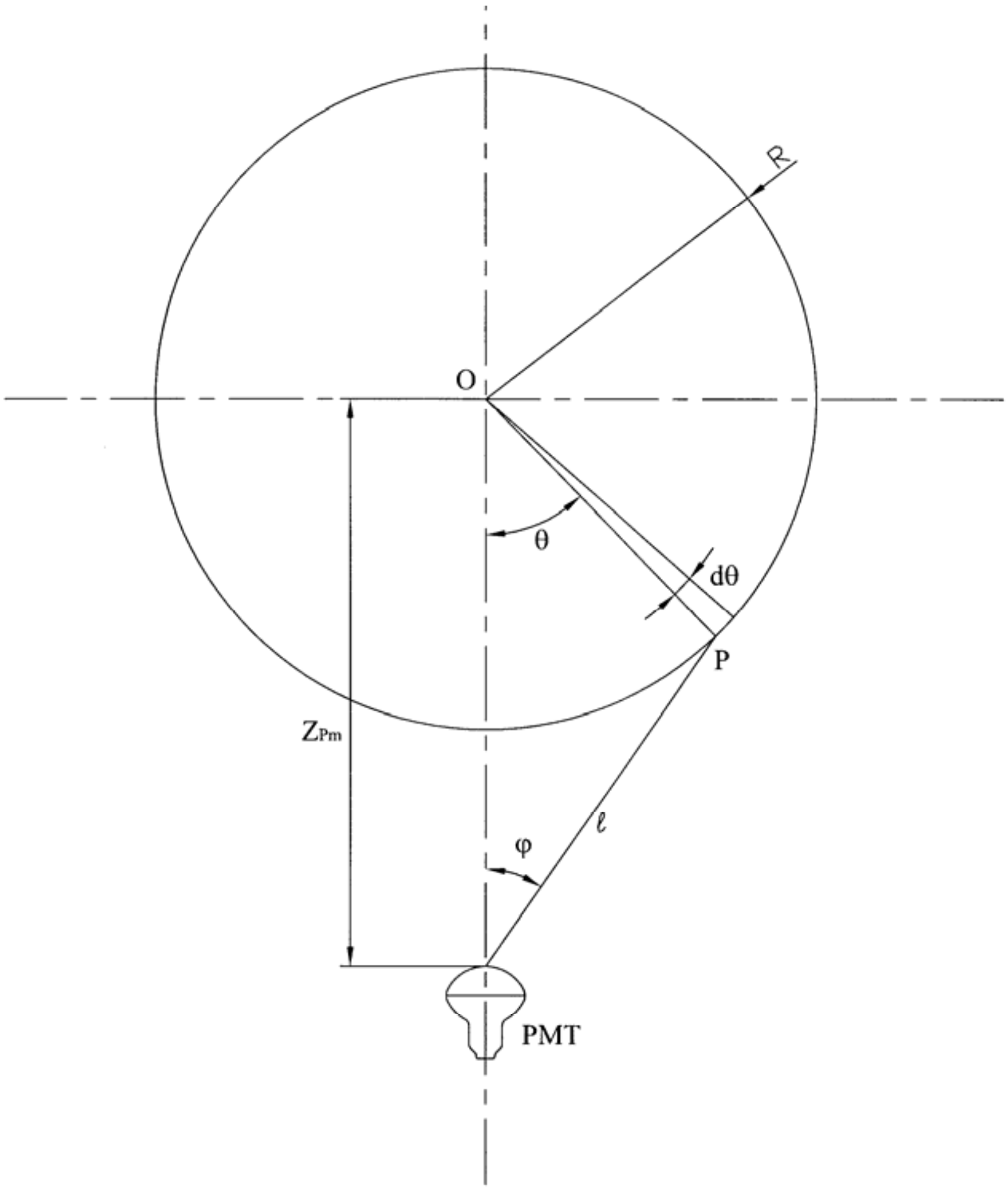

*Fig. 4 - Geometrical derivation of the time of flight distribution detected by a PMT observing a spherical detector characterized by a uniform distribution of scintillation events only on its surface*

By differentiating both members of eq. (14) we get

$$\sin\theta d\theta = \frac{l}{RZ_{pm}} dl \quad (15).$$

From the same triangle we can get also $cos\varphi$

$$\cos\varphi = \frac{Z_{pm}^2 + l^2 - R^2}{2lZ_{pm}} \quad (16).$$

Introducing the relation (15) and (16) in the eq. (13) we get

$$p(l)dldS = \frac{1}{2}\frac{l}{RZ_{pm}}dl\frac{dS}{4\pi l^2}\frac{Z_{pm}^2 + l^2 - R^2}{2lZ_{pm}} \quad (17)$$

$$p(l)dldS = \frac{1}{16\pi}\frac{Z_{pm}^2 + l^2 - R^2}{Rl^2 Z_{pm}^2}dldS \quad (18)$$

and so finally

$$p(l) = \frac{1}{16\pi}\frac{Z_{pm}^2 + l^2 - R^2}{Rl^2 Z_{pm}^2}dS \quad (19)$$

where the meaning of the surface element $dS$ is as previously explained for eq. (9).
The relation (19), expressed in term of the time of flight, becomes

$$p(t) = \frac{1}{16\pi}\frac{Z_{pm}^2 + \left(\frac{c}{n}t_f\right)^2 - R^2}{R\left(\frac{c}{n}t_f\right)^2 Z_{pm}^2}\frac{c}{n}dS \quad (20)$$

As in the previous paragraph, Eq. (20) can also be modified to account for the attenuation effect as follows

$$p(t) = \frac{1}{16\pi}\frac{Z_{pm}^2 + \left(\frac{c}{n}t_f\right)^2 - R^2}{R\left(\frac{c}{n}t_f\right)^2 Z_{pm}^2}\frac{c}{n}e^{-\frac{\frac{c}{n}t_f}{l_{att}}}dS \quad (21).$$

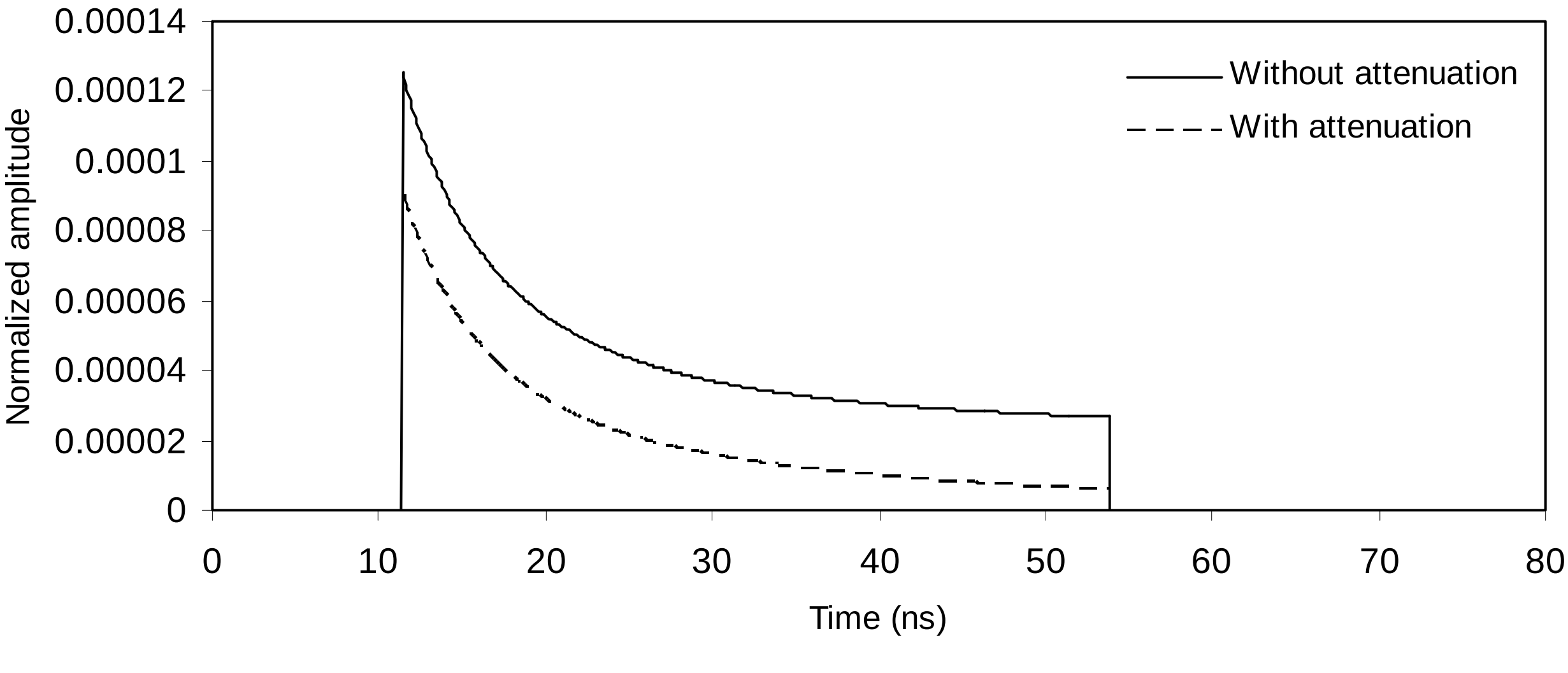

*Fig. 5 - Time of flight detected by an individual phototube as response to uniformly distributed surface events*

(A similar modification is valid for relation (19), too).The time distribution for unit surface (i.e. eq. (21) omitting the factor $dS$) is plotted in Fig. 5 for the geometrical parameters corresponding to the Borexino detector ($R$=4.25 and $Z_{pm}$=6.52); as for Fig. 3, the two cases without or with the attenuation ($l_{att}$=7 m) are considered.

## 6. Time distribution detected by a phototube for a uniform volume distribution with refraction at the boundary

An unusual situation which requires to modify the calculation in paragraph 4 is that in which the surrounding medium is different from the scintillator solvent, being typically water, instead. In this case there is refraction at the boundary which greatly complicates the evaluation of the time of flight at the phototube.

As depicted in Fig. 6, the ray emitted from the event point is bended at the boundary of the two mediums. Since the refraction index of the inner medium (scintillator) is lower than that of the outer medium (water) the direction of the ray is bended so to increase the angle with the radius of the sphere drawn through the incidence point. The evaluation of the time distribution at the phototube can be done numerically, since a closed form formula cannot be written. The numerical evaluation proceeds through two steps: the first is the determination of the optical path from the generic event site to the phototube, the second is the construction of the overall distribution from the individual path lengths.

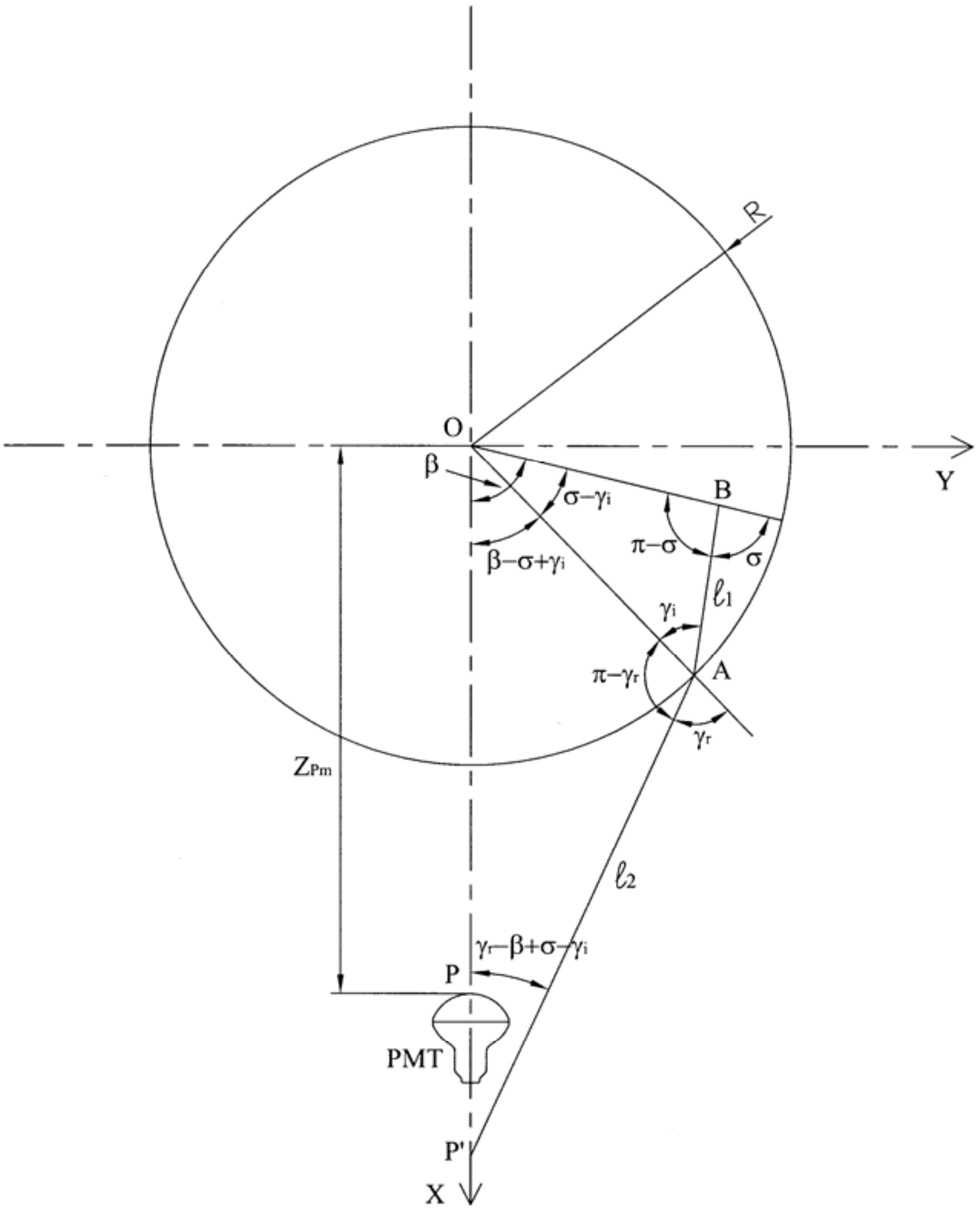

*Fig.6 - Geometrical derivation of the time of flight distribution detected by a PMT observing a spherical detector characterized by a uniform internal distribution of scintillation events and with light refraction at the boundary*

The construction in Fig. 6 has the purpose to illustrate the determination of the optical path. Given B the event point, the path traveled by the photon to arrive to the PMT is the sum of the two segments $l_1$ and $l_2$ shown in the figure. In the adopted reference coordinate system, B has the generic coordinates $x$ and $y$ and the abscissa of the PMT is $Z_{pm}$. The direction of the radius passing through B is specified by the angle $\beta$. The incident angle is denoted as $\gamma_i$, the refraction angle as $\gamma_r$, the angle formed by the photon with respect to the radius passing through B is denoted as $\sigma$. The straight distance from the point B to the PMT is obtained via the cosine theorem applied to the triangle OBP

$$BP = Z_{pm}^2 + OB^2 - 2OB \cdot Z_{pm} \cos\beta \tag{22}$$

In order to compute the right path $l_1+l_2$ bringing the photon from the event site to the PMT, an iterative procedure is adopted: many trial values of the angle $\sigma$ are tried (starting from 0) and for each of them the following values are computed :

$$l_1 = \frac{R}{\sin\sigma}\sin(\sigma - \gamma_i) \tag{23}$$

from the sine theorem applied to the triangle OBA, and

$$l_2 = \frac{R\sin(\beta - \sigma + \gamma_i)}{\sin(\gamma_r - \beta + \sigma - \gamma_i)} \tag{24}$$

from the sine theorem applied to the triangle OAP'.

The special values of $l_1$, $l_2$ for which they form together with BP' a triangle with vertexes B, A and P (i.e. when P and P' coincide) are the required correct segments of the optical path, and at this point the iteration is stopped. Again with reference to Fig. 6, it can be demonstrated that pre-requisite for the refracted ray to hit the phototube is that the condition $\alpha = \gamma_r - \beta + \sigma - \gamma_i > 0^\circ$ holds, where $\gamma_r - \beta + \sigma - \gamma_i$ is the angle formed by the path $l_2$ with the positive $x$ axis. In the opposite case $\alpha = \gamma_r - \beta + \sigma - \gamma_i < 0^\circ$ the ray $l_2$ would not hit the positive $x$ axis, but on the contrary it would be the prolongation on the opposite side of $l_2$ that would hit the negative $x$ axis (forming the angle $-\alpha$), thus giving origin to a construction in which the PMT is surely not hit by the light.

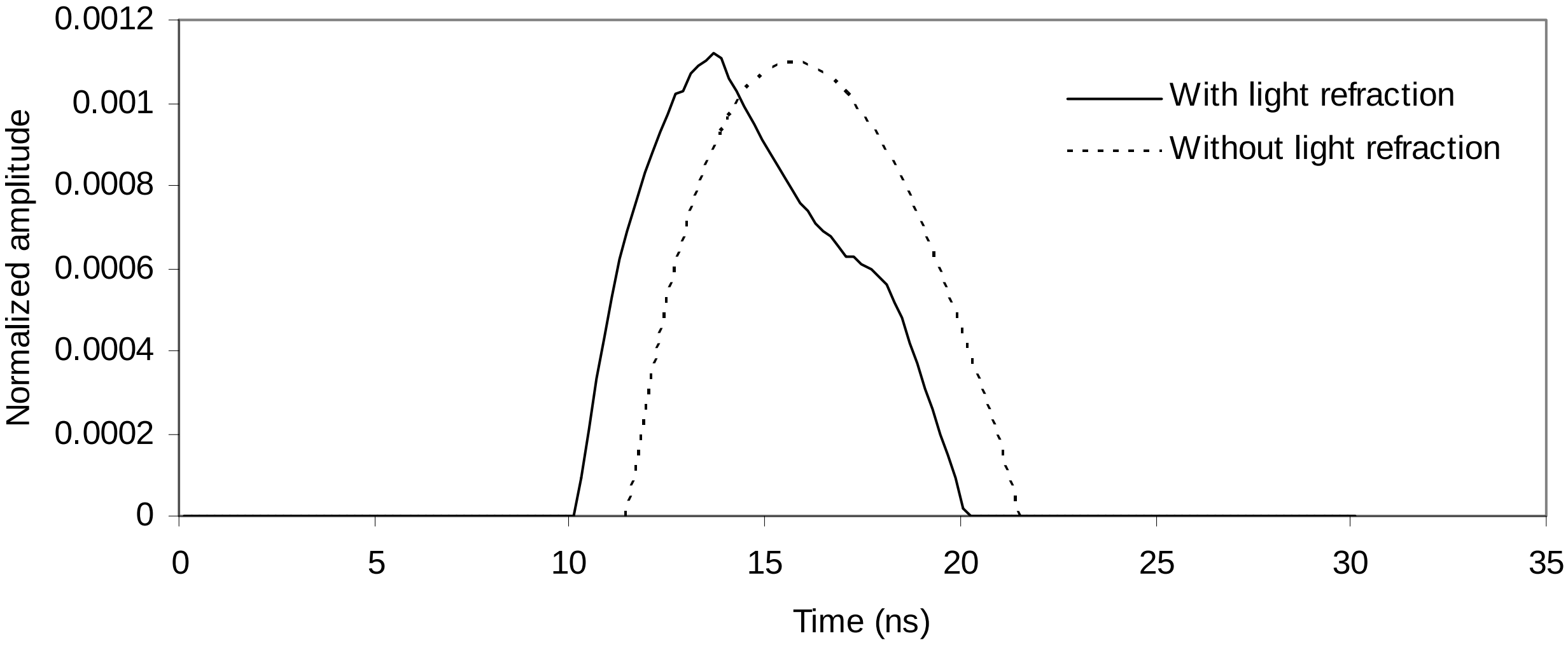

*Fig. 7 - Time of flight distribution induced on an observing phototube by a uniform distribution of events in the scintillation vessel in the case of two different medium (scintillator and water). The example has been computed for the parameters characterizing the CTF detector. For reference purpose the distribution in case of absence of refraction (dashed curve) is also reported*

The light paths from all the points of the sphere can be properly grouped to form the overall distribution of the time of flight to the PMT. Taking into account the spherical symmetry of the problem, it is enough to consider a grid of calculation points in a semicircle of the sphere itself:

assuming a certain binning for the time of flight distribution to be derived (for example 0.2 ns) each path corresponding to a point of the grid is converted into a bin number, and to the content of the bin so identified it is added a factor which represents the weight of the calculated path. Since each point of generic coordinates *x*, *y* actually stands for an entire circular corona of radius *y* (in the sense that all the points comprised in this corona produce the same path to the phototube), the first factor in the weight is simply the volume of such a corona, i.e. $2\pi y\Delta x\Delta y$, being $\Delta x$ and $\Delta y$ the widths of the cells of the grid.

The second factor in the weight gives the probability that the photon actually impinges the detecting phototube, and it is thus expressed as $\frac{dS}{4\pi(l_1+l_2)^2}\cos(\gamma_r-\beta+\sigma-\gamma_i)$.

Therefore the weight is globally given by

$$W = 2\pi y\Delta x\Delta y\frac{dS}{4\pi(l_1+l_2)^2}\cos(\gamma_r-\beta+\sigma-\gamma_i) \quad (25).$$

In practice $dS$ is omitted, as above, and the calculation are referred to the unit area of the detecting sphere of the phototubes.

A detector in which the refraction effect is important is the Counting Test Facility (CTF) [7], prototype of the above mentioned Borexino detector. In CTF the inner vessel ($R$=1) containing the scintillator with refraction index equal to 1.5, is surrounded by water (refraction index 1.33). The output of the determination of the TOF factor with these parameters (and considering that in CTF the phototubes are located at $Z_{pm}$=3.3 from the centre) is shown in Fig. 7, where for comparison it is reported also the time of flight distribution that would result if the two mediums were the same (with $n_s$=1.5) everywhere. The shift of the two curves is due to the systematically longer optical path in the latter situation, due to the assumption of having in the whole detector the medium with the higher refraction index. Besides this, in the refraction case the curve appears also distorted with a sort of "missing portion". This peculiar shape is actually linked to the presence of points in the vessel which are missed by the phototube, in the sense that the light rays emerging from them cannot find a path suitable to reach the observing phototubes because of the refraction effect, which can also give rise to a light trapping at the border. For completeness, in Fig. 8 the map of the obscured points is reported.

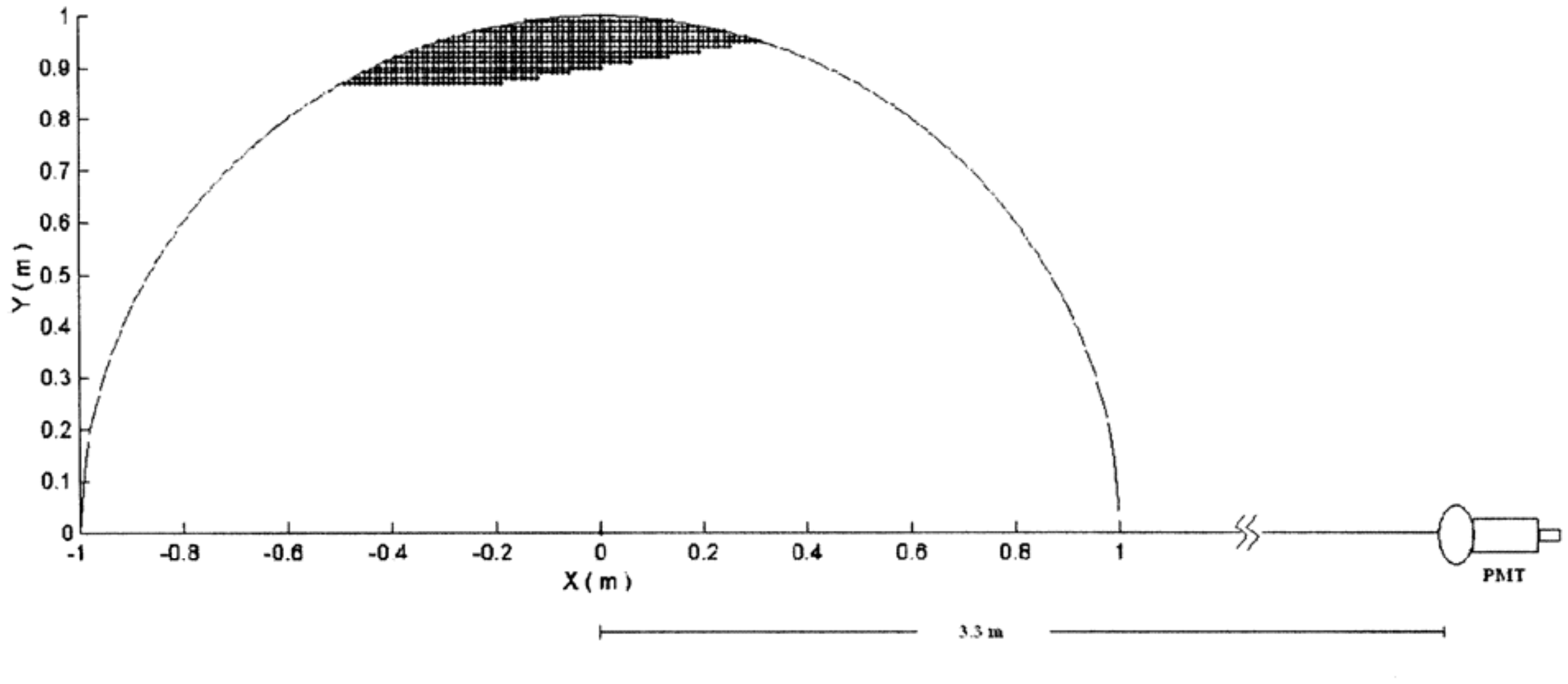

*Fig. 8 – Map of the points inside the CTF vessel for which an observing PMT is blind*

In this respect it is interesting to point out that, if the distribution is referred to the whole detecting surface through an integration which is equivalent to a multiplication by the area of such a surface, the re-scaled distribution will have an integral, less than 1, which gives the fraction of the sphere which is not obscured. In the present CTF example this fraction amounts to 91.3%.

If required, in the procedure described in this paragraph it can be included, as well, the attenuation effect, by introducing in the weight (25) the appropriate exponential factors containing $l_1$ and $l_2$, depending upon the actual physical situation. In the CTF example this correction has not been included for the twofold reason that the attenuation outside is negligible, because the medium is water, and inside is negligible also, because of the limited size of the detector.

Finally, it must be mentioned that as a cross-check, the procedure adopted in this paragraph has been applied to the case of same index of refraction everywhere, obtaining a distribution coinciding with the correct curve described by eq. (12) (without the attenuation factor).

**7. Time profile of single events**

Up to now we have considered how a single phototube detects the time distribution of the photons coming from a uniform (or surface) distribution of events in the scintillator containment vessel. An alternative distribution of interest is how the photons from a specific event site are distributed in time. Essentially, this is the opposite situation: instead of evaluating the distribution induced on a single PMT by an ensemble of events, we are now interested to the time distribution on all the PMT's of the detection times of the photons induced by a single interaction site.

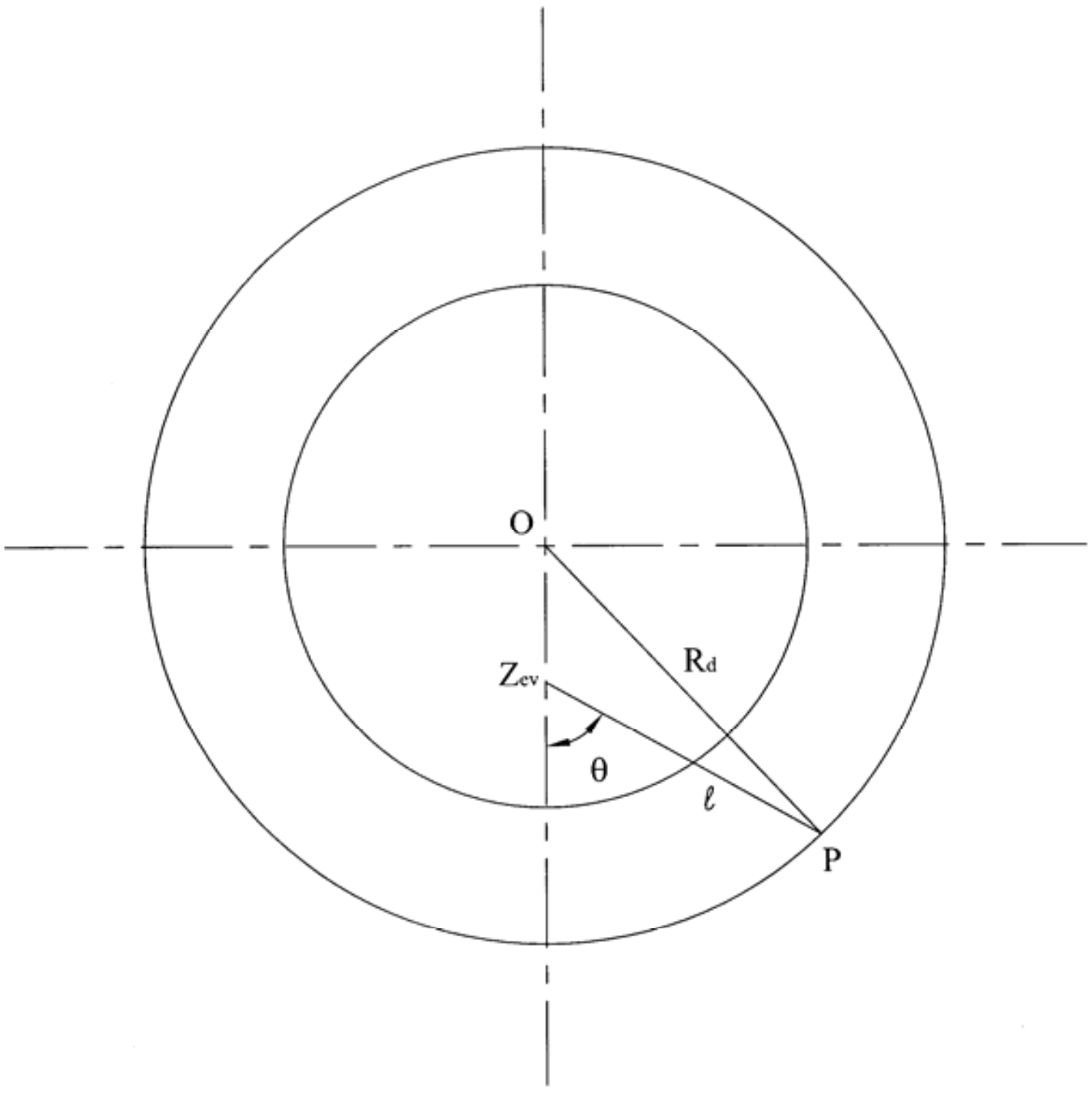

*Fig. 9 – Geometry for the derivation of the time of flight distribution recorded by a detecting sphere of radius $R_d$ as response to a localized interaction in the scintillator*

Again we neglect the absorption and re-emission effect and limit ourselves to consider the "zero order" effects, due to the geometry and to the attenuation process. The calculation in the next two sub-paragraphs takes into account two different situations: either same scintillation medium everywhere or two different mediums inside and outside the vessel. In both cases the first step of the calculation is the determination of the "time of flight" factor (TOF), which in this situation is the distribution of the arrival times to the detecting sphere (i.e. the lattice of phototubes) of the photons as if released in an ideal instantaneous scintillation process; the second step is the convolution of the TOF term with the intrinsic scintillation decay curve and the response of the PMT in order to obtain the overall time distribution of the detected photoelectrons.

*7.1 Same index of refraction*

With reference to Fig. 9, where the emission point is denoted with $Z_{ev}$, the generic arrival point with $P$, and the path traveled by the photon with $l$, if there is no absorption the time of flight distribution of the photons arriving to the detecting sphere of radius $R_d$ is determined by pure geometrical arguments.

Indeed it is obvious that the distribution of the time of flight is dictated by the distribution of the solid angle $\Omega$. Specifically, the photons emitted in the infinitesimal solid angle $d\Omega$ will be characterized by the same time of flight. Hence we have to transform the probability that a photon is emitted between $\Omega$ and $\Omega+d\Omega$, in the probability that its time of flight is comprised between $t_f$ and $t_f+dt_f$.

To do this we have to use the relationship between $\Omega$ and $\theta$, the vertex angle of the cone which subtends $\Omega$

$$\Omega = 2\pi(1-\cos\theta) \tag{26}$$

from which we get

$$d\Omega = 2\pi \sin\theta d\theta \qquad (27).$$

Since the probability that a photon is emitted between $\Omega$ and $\Omega+d\Omega$, $p(\Omega)d\Omega$, is equal to $\frac{1}{4\pi}d\Omega$ (the emission is isotropic), by multiplying both terms of eq. (27) by $1/(4\pi)$, we get

$$\frac{1}{4\pi}d\Omega = \frac{1}{2}\sin\theta d\theta\,. \qquad (28).$$

By applying the cosine theorem to the triangle $OZ_{ev}P$ we obtain

$$\cos\theta = \frac{R_d^2 - Z_{ev}^2 - l^2}{2Z_{ev}l}\,. \qquad (29).$$

By differentiating the first member of eq. (29) with respect to $\theta$, and the second to $l$, we get

$$-\sin\theta d\theta = \frac{-2l\cdot 2Z_{ev}l - 2R_d^2 Z_{ev} + 2Z_{ev}^3 + 2Z_{ev}l^2}{4Z_{ev}^2 l^2}dl \tag{30}$$

$$-\sin\theta d\theta = \frac{-2l^2 Z_{ev} - 2R_d^2 Z_{ev} + 2Z_{ev}^3}{4Z_{ev}^2 l^2}dl \tag{31}$$

$$\sin\theta d\theta = \frac{1}{2Z_{ev}}\left(1+\frac{R_d^2 - Z_{ev}^2}{l^2}\right)dl \tag{32}$$

or dividing both members by 2

$$\frac{1}{2}\sin\theta d\theta = \frac{1}{4Z_{ev}}\left(1+\frac{R_d^2 - Z_{ev}^2}{l^2}\right)dl \qquad (33).$$

Through the comparison of eq. (28) with eq. (33) we realize that the probability density function of the flight path of the not absorbed photons is

$$p(l) = \frac{1}{4Z_{ev}}\left(1 + \frac{R_d^2 - Z_{ev}^2}{l^2}\right) \qquad (34).$$

We can transform the last expression (34) in probability density function of the time of flight by the usual transformation $l=(c/n)t_f$, thus obtaining

$$p(t_f) = \frac{1}{4Z_{ev}}\left(1 + \frac{R_d^2 - Z_{ev}^2}{\left(\frac{c}{n}t_f\right)^2}\right)\frac{c}{n} \qquad (35).$$

Eq. (35) is valid on a limited interval of values of the variable $t_f$; indeed the minimum value of the time of flight corresponds to the path $(R_d\text{-}Z_{ev})$ and the maximum to $(R_d+Z_{ev})$. The eq. (35) hence is valid for

$$(R_d - Z_{ev})\frac{n}{c} \le t_f \le (R_d + Z_{ev})\frac{n}{c}.$$

It can be easily checked that, as expected, the integral of eq. (34) over $l$ (or equivalently the integral of relation (35) over $t_f$) is equal to 1.

To account for the attenuation effect, the distribution (35) must be corrected, as done for the distributions in paragraphs 4 and 5, introducing an exponential factor, therefore obtaining

$$p(t_f) = \frac{1}{4Z_{ev}}\left(1 + \frac{R_d^2 - Z_{ev}^2}{\left(\frac{c}{n}t_f\right)^2}\right)\frac{c}{n}e^{-\frac{\frac{c}{n}t_f}{l_{att}}} \qquad (36).$$

The medium outside the vessel is equal to the medium inside the vessel itself, leading to the same attenuation effect everywhere.

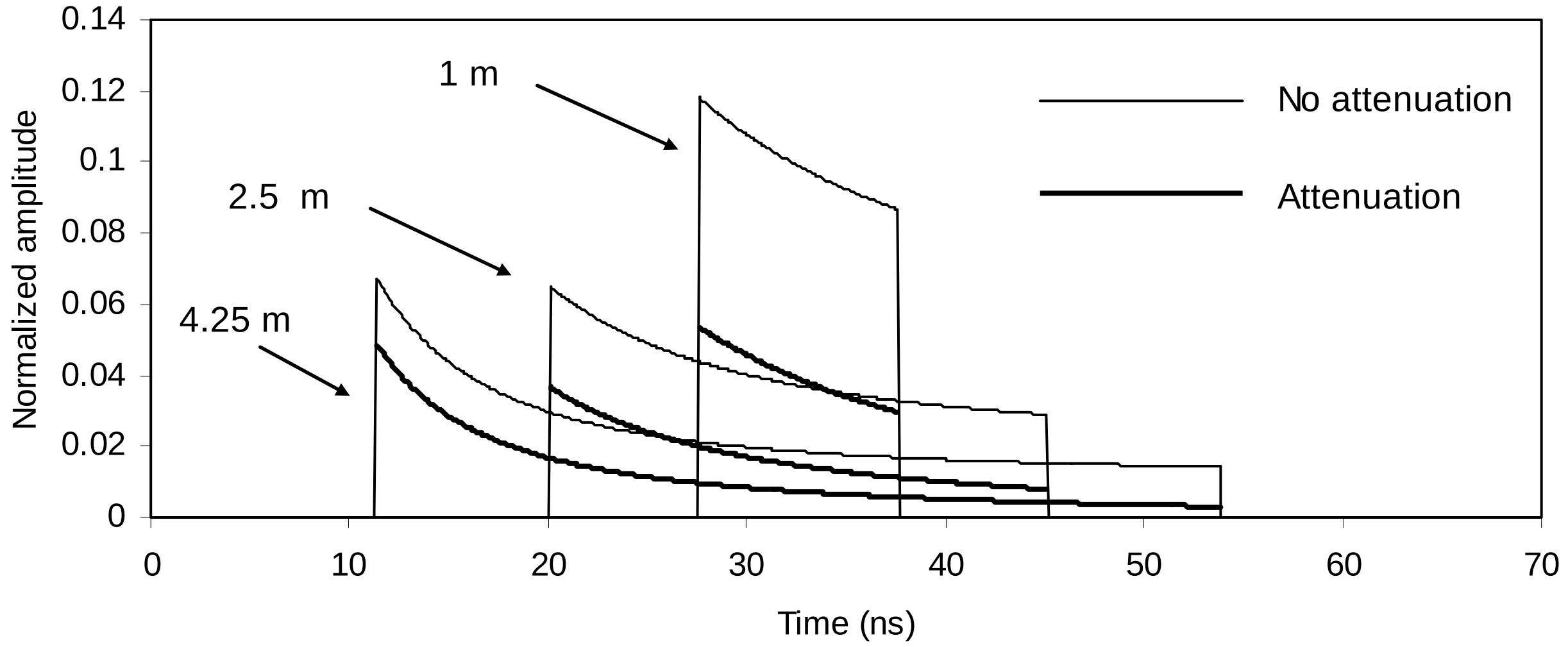

*Fig. 10 – Time of flight distributions of the photons released by events located respectively at 1m, 2.5 m and 4.25 m from the centre of the Borexino detector, with or without the inclusion of the attenuation effect*

The TOF factors (35) and (36), with reference to the geometrical parameters of the Borexino detector, are plotted in Fig. 10 for 3 different positions in the vessel (1, 2.5, 4.25 m): it is evident that the broadening effect due to the geometry, obviously more pronounced for locations of the events closer to the boundary, is accompanied by a peculiar peaking of the time distributions at shorter values. The curves including the attenuation effect, evaluated assuming an attenuation length of 7 m, feature a very similar shapes to those unaffected by the attenuation; their integral, which is less than 1, is physically meaningful and gives the probability of a photon to reach the detecting surface without being absorbed.

*7.2 Case of index of refraction discontinuity*

With reference to Fig. 11, in case of two different mediums the overall time distribution of the photons impinging upon the detecting sphere (i.e. the sphere of the phototubes) is obtained numerically following a procedure similar to that used in paragraph 6; see also the independent derivation described in [24]. Preliminarily it is evaluated, for each angle $\theta$, the light path $l_1+l_2$. Practically, the evaluation is carried out for a fine grid of $\theta$ values, with a step $\Delta\theta$ which can be done as small as desired. The flight time for each $\theta$ is then $\frac{n_s}{c}l_1 + \frac{n_w}{c}l_2$; upon defining a suitable binning for the desired distribution, for example 0.2 ns, the flight time is used to identify the corresponding bin. To that bin it is thus added the proper weight corresponding to the computed time of flight, which is equal to

$$W = e^{-\frac{l_1(\theta)}{l_{att}}} \frac{1}{2}\sin\theta\Delta\theta \qquad (37)$$

where the exponential term represents the probability of the photon of being not absorbed within the vessel (assuming that outside there is water, whose attenuation effect is hence negligible).

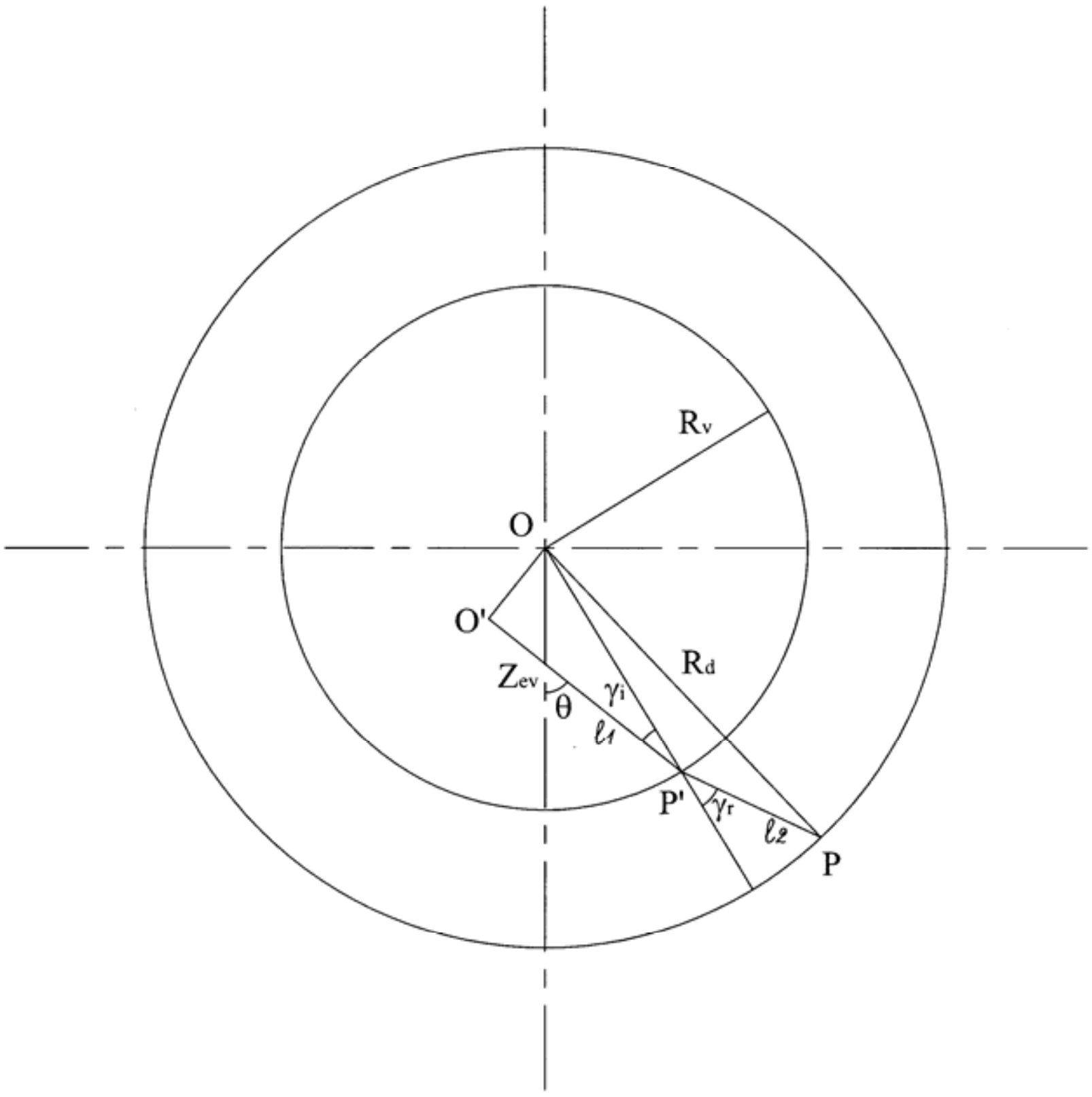

*Fig.11 – Geometry for the derivation of the time of flight distribution recorded by a detecting sphere of radius $R_d$ as response to a localized interaction in the scintillator, with light refraction at the border of the vessel*

According to the Fig. 11, the relevant formulas are:

$$\gamma_i = \arcsin\left(\frac{Z_{ev}}{r_v}\sin\theta\right) \tag{38}$$

$$\gamma_r = \arcsin\left(\frac{n_s}{n_w}\sin\gamma_i\right) \tag{39}$$

$$l_1 = -Z_{ev}\cos\theta + \sqrt{r_v^2 - (Z_{ev}\sin\theta)^2} \tag{40}$$

(obtained from the triangles OO’P’ and OO’$Z_{ev}$)

$$l_2 = -r_v\cos\gamma_r + \sqrt{R_d^2 - (r_v\sin\gamma_r)^2} \tag{41}$$

(obtained from the cosine theorem applied to the triangle OP’P).

The TOF distribution, evaluated according the prescription just illustrated, is plotted in Fig. 12 for 5 different radial positions in the CTF vessel (due to the limited size of the detector, the attenuation factor in this case has been dropped from the weight). It is interesting to note the effect due to the total internal reflection for the points close to the border: essentially there is a central gap in the time distribution due to the missing trapped photons. For display purpose the curves are all normalized to unit area. While this normalization is the right one for the curves not affected by the light trapping, this is not true for the last two curves exhibiting the trapping effect. It must be pointed out that the calculation as carried out provides in principle for them the proper normalization, thus leading to an integral that gives correctly the fraction of not trapped photons.

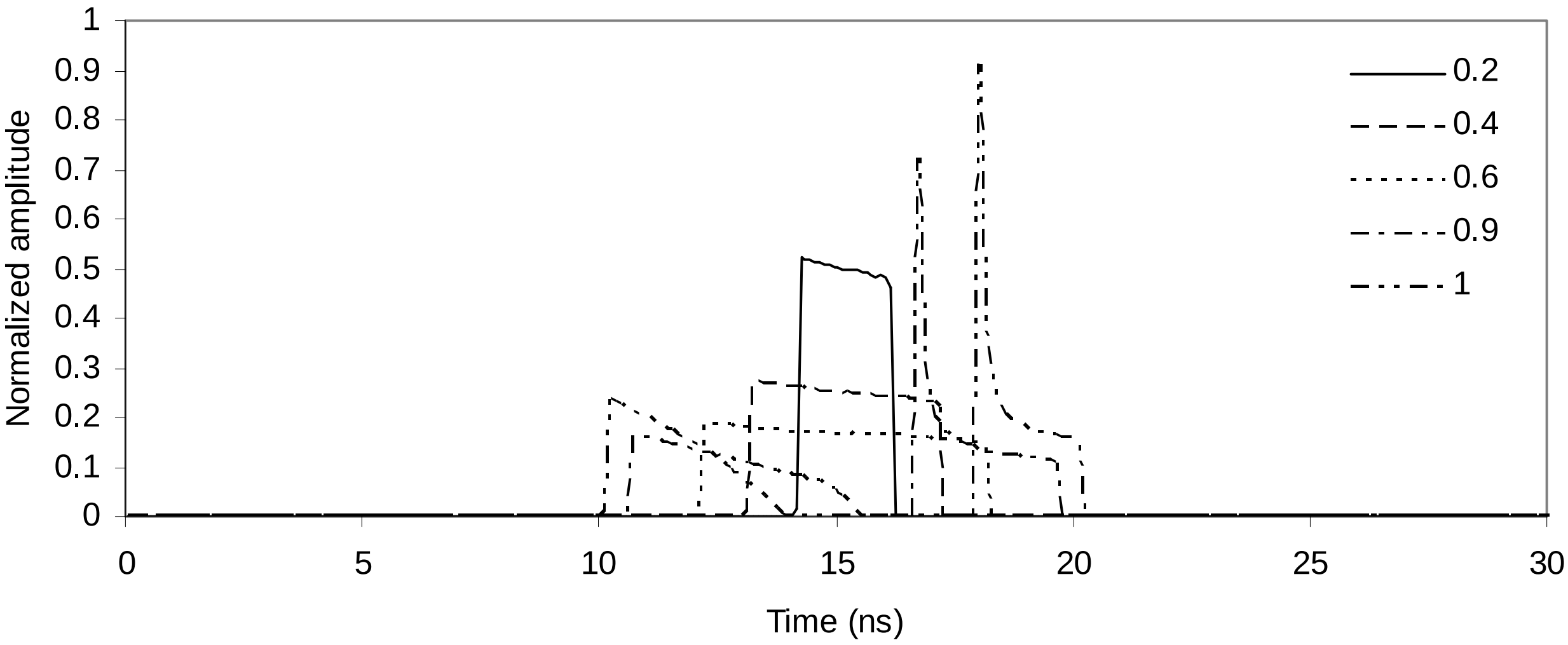

*Fig. 12 – Time of flight distributions of the photons released by events located, respectively, at 0.2 m, 0.4 m, 0.6 m, 0.9 m and 1 m from the centre of the CTF detector; it should be noted the gap in the distributions affected by the light trapping, as well as the very peaked characteristic of the second part of these distributions*

Another peculiarity of the trapped curves is the sharp peak that they feature at the beginning of their second part. While this characteristic stems from details of the mathematical calculation that are difficult to predict “a-priori”, its presence can be qualitatively understood as the effect of the abrupt transition from the “dark” to the “non-dark” region. To shed more light on this intriguing feature, it may be interesting in this case to carry out explicitly the calculation to go from the TOF distributions of the photons to the actual detected distributions of the photoelectrons, according to the sequence of convolution operations outlined in paragraph 3.

The results of the convolution of the TOF terms in Fig. 12 with the intrinsic scintillation light profile and the phototube timing response (we remind that we adopt for the sake of the example the quantities valid for the Borexino project) are reported in Fig. 13. Also in this case the curves are all normalized to 1 for the purpose of comparison. The interesting result is that for the trapping related curves the double peak feature of the photon time of flight is maintained also in the final photoelectrons distributions, being the second peak more prominent than the first.

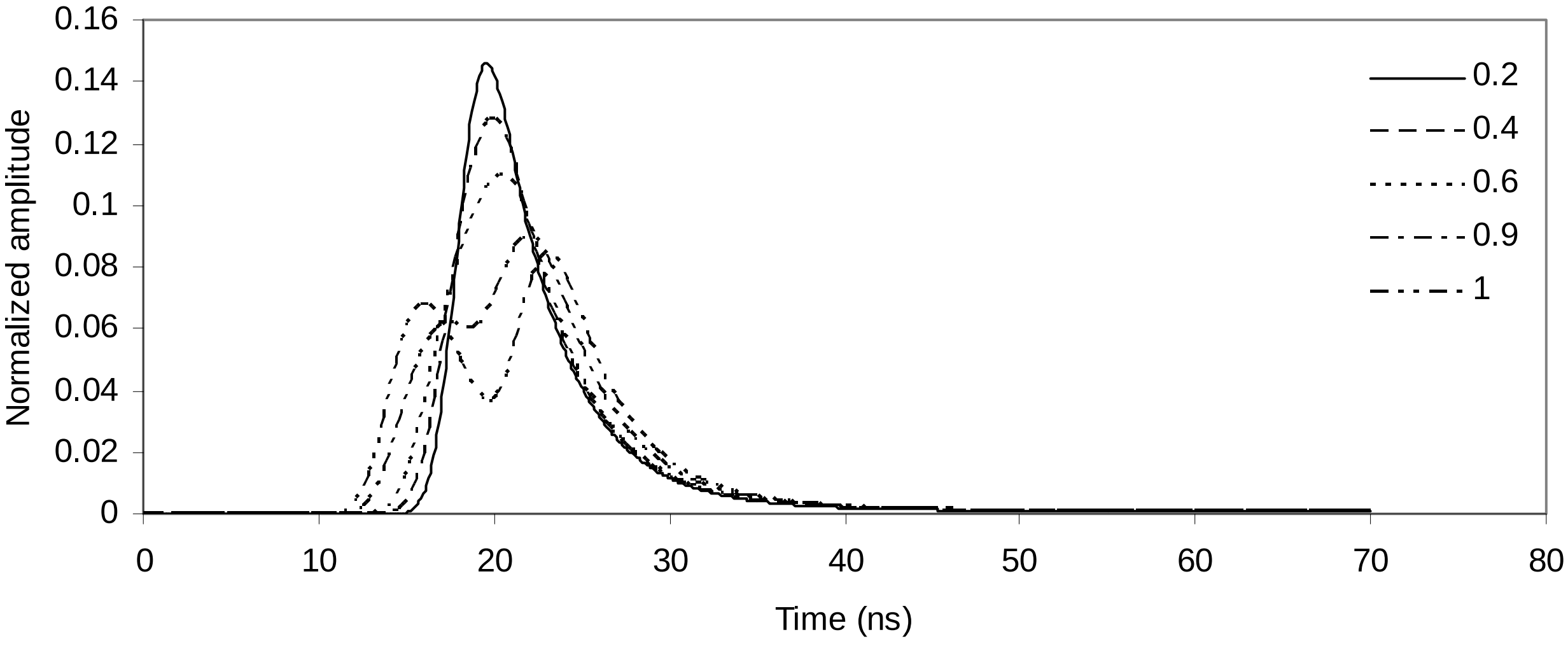

*Fig. 13 – Detected photoelectron distributions induced by events located, respectively, at 0.2 m, 0.4 m, 0.6 m, 0.9 m and 1 m from the centre of the CTF detector; the distributions affected by the light trapping phenomenon exhibit a double peaked structure reflecting that of the original photon time of flight, being the second peak more marked than the first*

This result can be compared, at least qualitatively, with the data gathered in CTF. In particular, the light trapping effect in CTF has been specifically investigated in [20], where it was clearly shown that events originating from locations close to the vessel (either intrinsic events or events generated by a calibration source) exhibit the double peaked structure retrieved by the present calculation, with the remarkable confirmation that also in the real data the second peak is more pronounced than the first, in agreement with the prediction of the model.

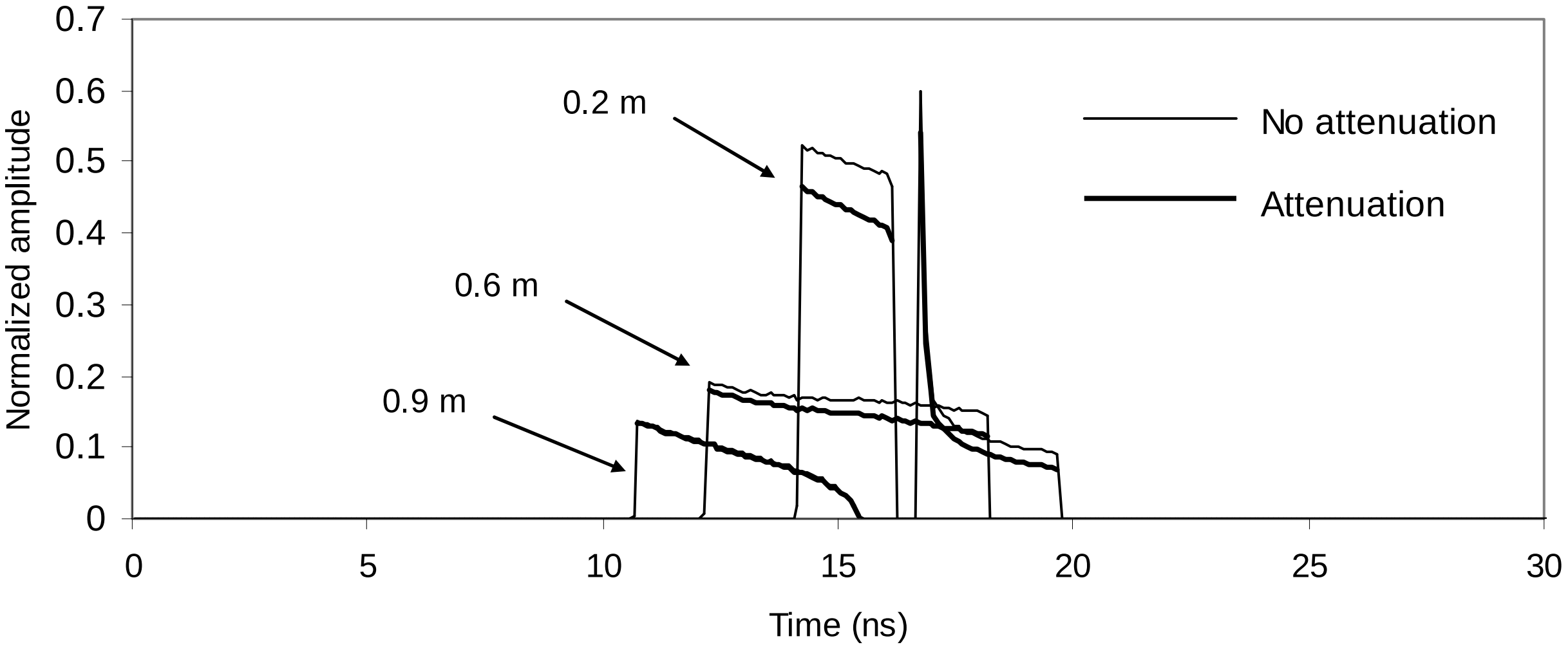

*Fig. 14 – Time of flight distributions of the photons released by events located, respectively, at 0.2 m, 0.6 m and 0.9 m from the centre of the CTF detector, with or without the inclusion of the attenuation effect*

For the purpose of completeness the TOF distribution is reported in Fig. 14 taking into account both the attenuation factor and the proper normalization due to the light trapping, in the three cases of event position at 0.2, 0.6 and 1 m. In correspondence of each position the curves relevant to the two different situations of attenuation included or not included are plotted. Obviously, for points well inside the vessel the only photon loss effect is that due to the attenuation: for the 0.2 m position the fraction of photons lost for the attenuation is 13.1 %, and at 0.6 m is 11.5 %. For points close to the boundary the loss of light due to the trapping is more important than the loss caused by the attenuation; for example for the position at 0.9 m the light trapping causes a loss in terms of photons of 17.2%, while the attenuation causes and additional loss of only 7%, essentially concentrated in the second part of the curve. As shown in the figure, the first part is unaffected by the attenuation since it is relevant to photons which fly across the scintillator for a short path.

*7. 3 Effects of a thick containment wall*

The calculations performed in this and in the previous paragraphs do not account explicitly for the effect of the thickness of the wall of the scintillator containment vessel. In practice this effect is modest, especially considering the small index of refraction difference between the scintillator and the materials suitable to be used for the vessel manufacturing. For example, while for a common scintillator mixture the refraction index is of order of 1.50-1.53, that of a typical material like nylon employed for the vessel construction [21] is 1.53. The minimal amount of the expected effect is justified also by the circumstance that in practical set-ups the nylon wall is extremely thin.

In this respect a slightly different situation will be that of the forthcoming SNO+ experiment, mentioned in the introduction. The containment vessel in this case (still the same used by the now ended SNO experiment) is made by thick acrylic panels of few centimeter thickness, which will separate the inner active scintillator volume from the external shielding water. Therefore, it can be interesting to extend the calculations of the previous section 7.2 to the scintillator-acrylic-water interface.

For this purpose it is enough to generalize the situation depicted in Fig. 11, by adding to the overall time of flight from the event site to the detecting sphere also the path traveled by the photon inside the wall (and considering, obviously, the refraction process at the entrance and exit of the wall itself). By denoting with $r_v$ and $R_v$ the inner and outer radius of the containment vessel, it can be easily shown by a simple extension of the procedure of the section 7.2 that the two paths traveled by the photons, respectively, within the vessel wall and in water (the path in the scintillator is clearly unchanged) are given by

$$l_2 = -r_v \cos\gamma_r + \sqrt{R_v^2 - (r_v \sin\gamma_r)^2} \qquad (42)$$

and

$$l_3 = -R_v \cos\gamma_r' + \sqrt{R_d^2 - \left(R_v \sin\gamma_r'\right)^2} \qquad (43),$$

where

$$\gamma_r = \arcsin\left(\frac{n_s}{n_a}\sin\gamma_i\right) \qquad (44)$$

is the refraction angle at the scintillator-wall interface, and

$$\gamma_r' = \arcsin\left(\frac{n_s r_v}{n_w R_v}\sin\gamma_i\right) \qquad (45)$$

is the refraction angle at the wall-water interface.

Assuming a scintillation point in contact with the vessel, for which thus the optical modification due to finite wall thickness is maximum, in Fig. 15 we show the corresponding time distributions both with and without the wall effect for a detector of the same dimension of CTF, but surrounded by an acrylic vessel 5 cm thick. The change induced to the time distribution in this example is quite evident, since we choose to enhance the difference an unrealistic too large wall depth. In practice, a thickness of the order of several cm is appropriate, from the engineering point of view, for much larger vessels, like the 12 m diameter SNO vessel: with these realistic dimensions it can be easily verified that the resulting change of the time distribution is even difficult to appreciate.

In this framework it can be worthwhile to add that another factor to account for a more accurate description of the interface effect is the transmission coefficient at the boundary, which would simply play the role of a further multiplicative factor in the relation (37).

Finally, we conclude this digression by outlining that a similar procedure to include the effect of a thick wall can be used to properly modify, if needed, the time distributions detected by a single observing phototube described in the previous paragraphs.

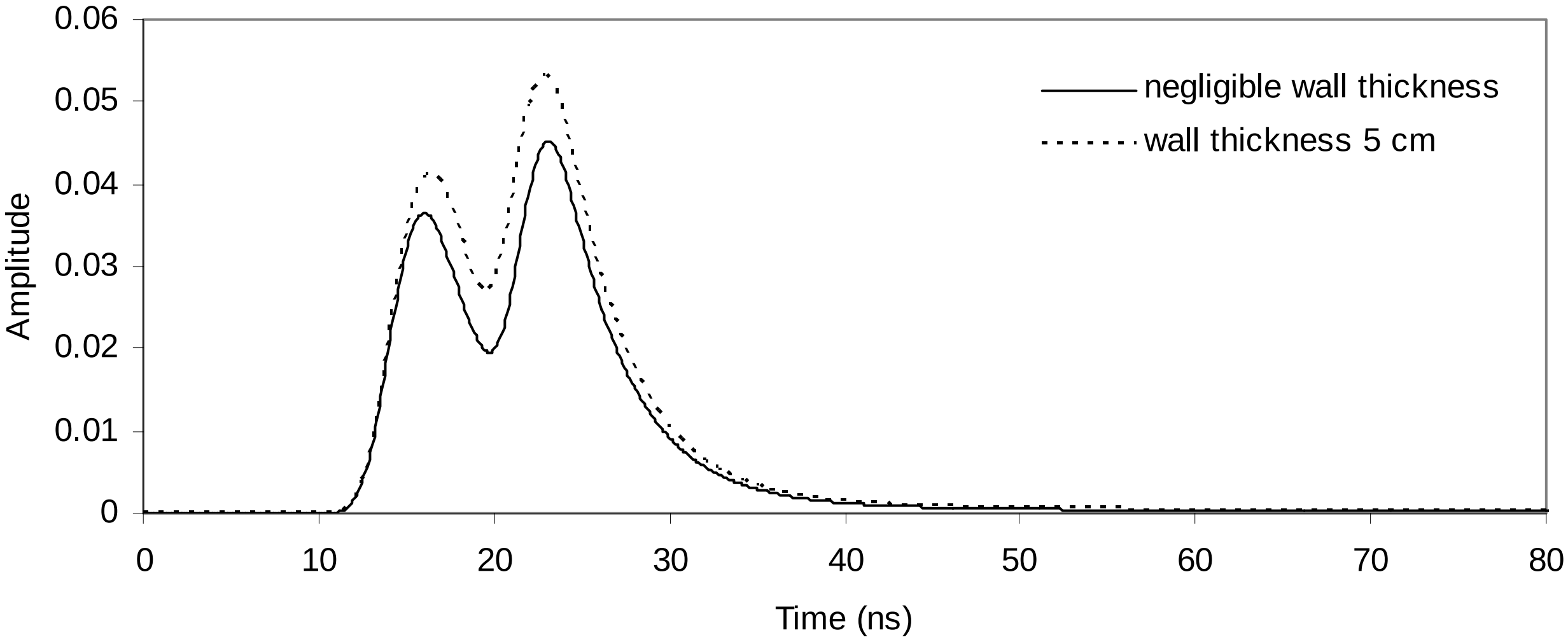

*Fig. 15 – – Detected photoelectron distributions induced by events located in contact with the vessel wall at 1 m from the centre of the CTF detector, respectively, with and without the inclusion of the effect of the wall thickness. The wall has been chosen on very thick (5 cm) to enhance the amount of the effect for the purpose of this example*

## 8. Trigger fluctuation

As explained in paragraph 2, the trigger condition in a large scintillation counter is obtained by requiring the firing of a certain number $N_F$ of PMT's in a suitable time window chosen on the basis of the detector extension. Obviously, with respect to the ideal 0 time of the actual occurrence of the event, the trigger time is affected by a fluctuation dictated by the statistics of the coincidence process. In the evaluation of the time sequence of the photoelectrons produced in a single event, this amount simply to an overall common shift of the time base; instead in the evaluation of the global time distribution detected by a single photomultiplier for a uniform volume or surface distributions of events this effect has an overall distortion impact due to the fact that the time measurements which are added together to create the global distribution are not referred to a same fixed 0 time but are referred to a fluctuating, and hence different on an event by event basis, trigger time. This implies, in particular, that the actual measured TOF distribution is the convolution of the trigger time distribution with the TOF ideal factor determined in the previous paragraphs.

It is, thus, of interest to find a way to compute the trigger fluctuation. This can be done, with a good degree of approximation, profiting of the known formulas describing the photon statistics in scintillation counters [22] [23]. Specifically, there are two applicable formulas in case that one considers events comprising either a fixed or a fluctuating numbers of photoelectrons. In the former case, given the probability density function of the photon production in the scintillator *p(t)*, then the probability of the time of production of the $i_{th}$ photon out of *n* photons exactly originated in a scintillation event is

$$p_i(t/n) = \frac{n!}{(i-1)!(n-i)!}[1-F(t)]^{(n-i)}[F(t)]^{i-1} p(t) \qquad (46)$$

where *F(t)* is defined as

$$F(t) = \int_0^t p(\lambda) d\lambda \qquad (47).$$

In the latter case, considering Poisson fluctuating events, the applicable formula is

$$p_i(t,n) = \frac{n^i e^{-nF(t)}[F(t)]^{i-1} p(t)}{(i-1)!} \qquad (48).$$

In the original contest in which they are derived, these formulas are related to the photon timing in a scintillator characterized by the scintillation light profile *p(t).* Since in the present case we are interested to the detection time of the photoelectrons as delivered by the phototubes, then the *p(t)* functions to be considered are the time profiles for single events as derived in paragraph 7: in fact, it is on the sequence of photoelectrons that obey to those curves (we remind that they amount to the convolution of the intrinsic scintillation light, the PMT response and the TOF factor, with or without the attenuation term) which is applied the coincidence criterion for the trigger derivation.

It can be instructive to evaluate for some special cases of number of detected photoelectrons, thus exploiting the formula (46), and for few points inside the scintillation vessel, the resulting trigger fluctuation. For example, considering a detector like Borexino, in which one can expect of the order of 20 PMT's in coincidence over a time window of 50 ns as trigger condition, it can be assumed as trigger distribution that of the detection time of the $20^{th}$ photoelectron, as derived from the formula (46). Clearly this is an approximation which is valid if the total number of photoelectrons is significantly high, since it is equivalent to assume that the photoelectron number 20 occurs within the first 50 ns window. In case of a number of photoelectrons not much higher than the trigger condition, it can happen that either in a substantial fraction of events the trigger condition is not met or that it is met, but not in the first 50 ns window.

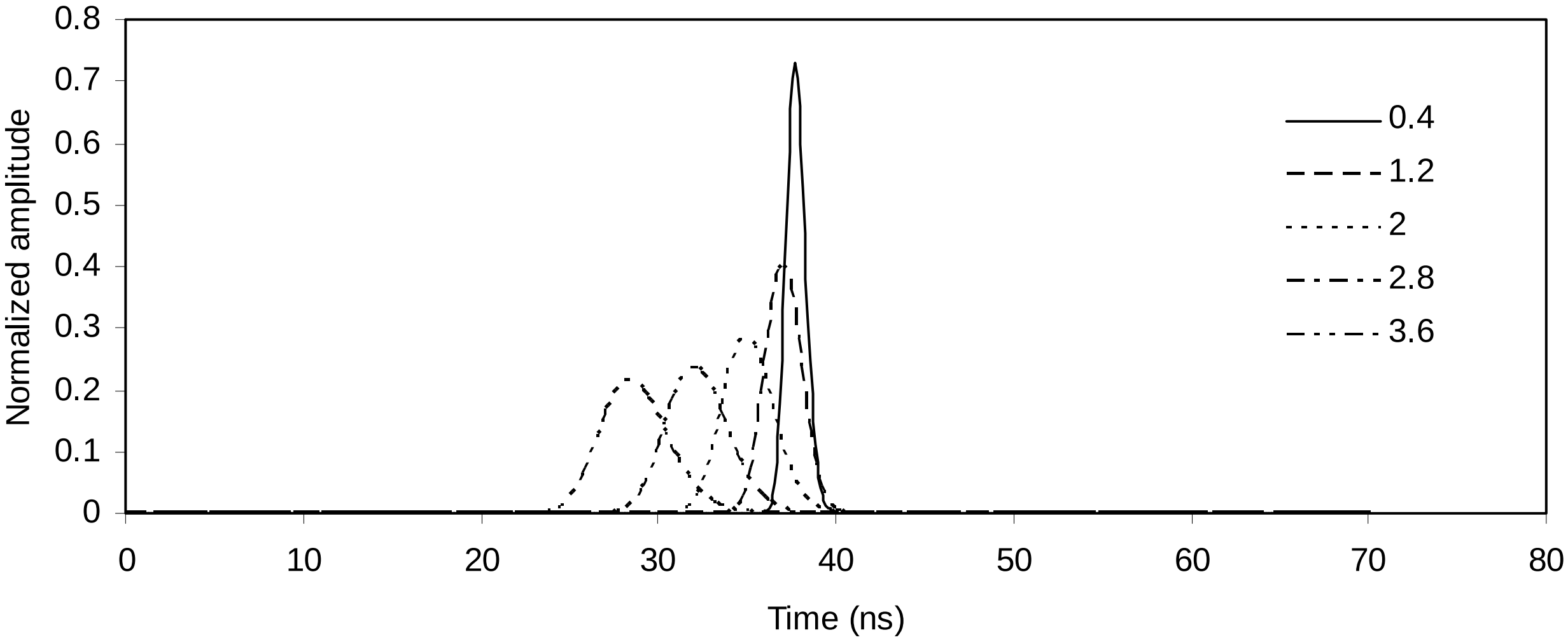

*Fig. 16 – Trigger time distributions in Borexino for events located at five different radial positions (0.4, 1.2, 2, 2.8, 3.6 m) and for 50 photoelectrons. The assumed trigger condition is 20 PMT's concurrently firing in a 50 ns time window*

In this case the evaluation of the trigger time could be done, besides obviously through a Monte Carlo calculation, following a sliding window approach which leads to a scan statistics calculation, which however is beyond the purpose of the present paper.

The trigger distributions in these conditions, for 50 detected photoelectrons and for 5 different radial position are reported in Fig 16. As intuitively expected, the dispersion of the trigger time is higher for events located at increasing radial distances from the center of the detector. Also, it appears that there is a shift on the mean value of the trigger time distributions depending upon the actual radial location of the events, with the events closer to the border causing obviously an early triggering with respect to those more central. By computing eq. (46) for increasing numbers of photoelectrons it appears that the trigger fluctuation curves shrink significantly, and also that their absolute shift with respect to the true 0 time decrease, since more photoelectrons induce obviously an early trigger occurrence.

## 9. Single coordinate distribution for uniform events

The sequence of the photoelectron detected times studied in the previous paragraphs can be used to infer “a-posteriori” the spatial location of the originating event. A thorough maximum likelihood procedure for this purpose has been studied for example in [13]. In this “inverse problem” framework the studies of the previous paragraphs can be extended deriving some interesting results concerning the expected spatial coordinates distribution of the events in some special cases. The main variable of interest is *r*, the radial coordinate of the location of the events, but it can be useful to show also the expected distribution of the individual *x*, *y* and *z* coordinates in the special case of uniform distribution of events. Let’s consider for this purpose that the probability for an event to be in the elementary volume *dxdydz*, in a spherical volume of radius *R*, is simply

$$p(x,y,z)dxdydz = \frac{1}{\frac{4}{3}\pi R^3} dxdydz \tag{49}$$

from which, by integrating over *y* and *z*, it stems the probability of the variable *x* to be between *x* and *x+dx*, i.e.

$$p(x)dx = \frac{dx}{\frac{4}{3}\pi R^3} \int_{-\sqrt{R^2-x^2}}^{\sqrt{R^2-x^2}} dy \int_{-\sqrt{R^2-x^2-y^2}}^{\sqrt{R^2-x^2-y^2}} dz \tag{50}$$

$$p(x) = \frac{1}{\frac{4}{3}\pi R^3} \int_{-\sqrt{R^2-x^2}}^{\sqrt{R^2-x^2}} dy \; 2\sqrt{R^2 - x^2 - y^2} \tag{51}$$

$$p(x) = \frac{3}{2\pi R^3} \int_{-\sqrt{R^2-x^2}}^{\sqrt{R^2-x^2}} \sqrt{R^2 - x^2 - y^2}\, dy \tag{52}$$

which by symmetry becomes

$$p(x) = \frac{3}{\pi R^3} \int_{0}^{\sqrt{R^2-x^2}} \sqrt{R^2 - x^2 - y^2}\, dy \qquad (53).$$

By putting $y = \sqrt{R^2 - x^2}\sin t$ and hence $dy = \sqrt{R^2 - x^2}\cos t dt$ eq. (53) becomes

$$p(x)=\frac{3}{\pi R^3}\int_0^{\arcsin\frac{y=\sqrt{R^2-x^2}}{\sqrt{R^2-x^2}}}\sqrt{R^2-x^2-(R^2-x^2)\sin^2 t}\cos t\sqrt{R^2-x^2}dt \qquad (54)$$

$$p(x)=\frac{3}{\pi R^3}\int_0^{\arcsin 1}\sqrt{(R^2-x^2)(1-\sin^2 t)}\cos t\sqrt{R^2-x^2}dt \qquad (55)$$

$$p(x)=\frac{3}{\pi R^3}\int_0^{\frac{\pi}{2}}(R^2-x^2)\cos^2 tdt \qquad (56)$$

$$p(x)=\frac{3(R^2-x^2)}{\pi R^3}\int_0^{\frac{\pi}{2}}\cos^2 tdt=\frac{3}{4R^3}(R^2-x^2) \qquad (57).$$

It can be easily checked that the relation (57) is correctly normalized to 1.

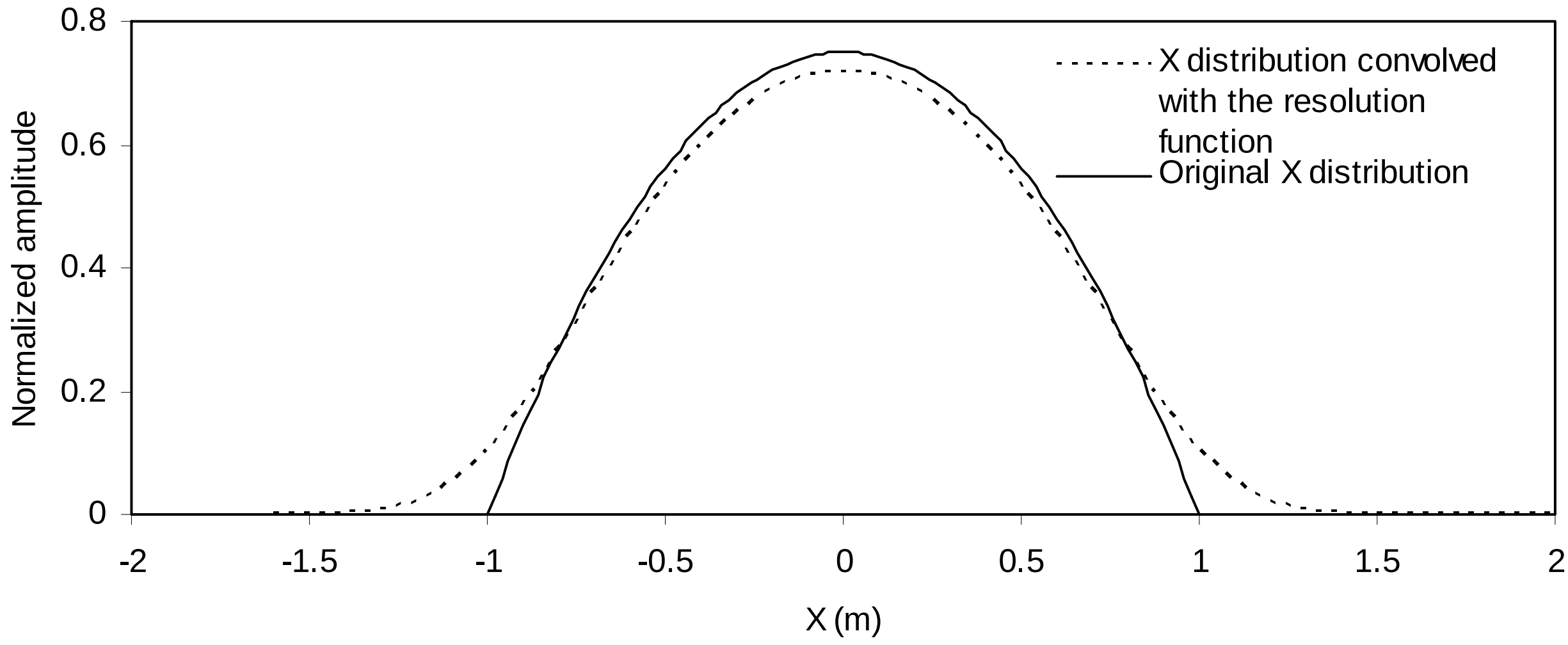

*Fig. 17 – Distribution of an individual spatial coordinate for events uniformly distributed in a vessel of 1m diameter, with and without the effect of the finite position resolution*

If the detection system (i.e. the lattice of the PMT's) is spatially uniform similar results are valid for the $y$ and $z$ coordinates, too.

So, if the detector is not affected by biases or systematic effects, a uniform distribution of events would result in parabolic distributions of the three spatial coordinates. Moreover, if the spread in the reconstruction process of the event coordinates is characterized (as it is usual) by a Gaussian profile, then the distribution of the reconstructed Cartesian coordinates is simply the convolution of the Gaussian resolution function with the parabolic function (57).

Eq. (57) and its convolution with the Gaussian spread are plotted in Fig. 17 for the case of a vessel of 1 m radius and assuming a resolution (sigma) of 20 cm. Interestingly, the convolution effect is to change marginally the bulk of the parabolic distribution, while adding two side tails whose extent is determined by the width of the resolution function.

**10. Reconstructed distribution of point-like events**

Again in this paragraph we do not consider how the event location is inferred starting from the sequence of the times of the detected photoelectrons, but we assume simply that such a procedure leads to estimates of the *x*, *y* and *z* coordinates which are all three characterized by a Gaussian uncertainty. What we want to show here is how these estimates combine to give the quantity of interest, i.e. the probability density function of the radial distance *r* of the event from the center of the detector. To this purpose we follow closely the derivation illustrated in [19]. Alternative approaches are also possible, like for example that reported in [24].

Let's consider the joint probability that the estimated coordinates are located around a generic point of coordinates *x*, *y*, and *z*. We can then write

$$p(x,y,z)dx,dy,dz = \frac{1}{\sigma_1\sigma_2\sigma_3\sqrt{(2\pi)^3}} e^{-\left(\frac{(x-x_0)^2}{2\sigma_1^2}+\frac{(y-y_0)^2}{2\sigma_2^2}+\frac{(z-z_0)^2}{2\sigma_3^2}\right)} dxdydz \qquad (58)$$

under the assumption that the three estimates are independent. Here $x_0$, $y_0$ and $z_0$ are the true, unknown coordinates of the event.

In polar coordinates we have, as usual $x=r\cos\varphi\cos\theta$, $y= r\cos\varphi\sin\theta$, $z=r\sin\varphi$ and

$dxdydz = r^2 \cos\varphi dr d\theta d\varphi$

from which eq. (58) becomes

$$p(x,y,z)dx,dy,dz = \frac{1}{\sigma_1\sigma_2\sigma_3\sqrt{(2\pi)^3}} e^{-\left(\frac{(r\cos\varphi\cos\theta-x_0)^2}{2\sigma_1^2}+\frac{(r\cos\varphi\sin\theta-y_0)^2}{2\sigma_2^2}+\frac{(r\sin\varphi-z_0)^2}{2\sigma_3^2}\right)} r^2\cos\varphi dr d\theta d\varphi$$

(59).

In order to obtain from the relation (59) the desired distribution in *r* it is enough to integrate the second member over $\theta$ and $\varphi$, so

$$p(r) = \frac{r^2}{\sigma_1\sigma_2\sigma_3\sqrt{(2\pi)^3}} \int_0^{2\pi} d\theta \int_{-\frac{\pi}{2}}^{\frac{\pi}{2}} e^{-\left(\frac{(r\cos\varphi\cos\theta-x_0)^2}{2\sigma_1^2}+\frac{(r\cos\varphi\sin\theta-y_0)^2}{2\sigma_2^2}+\frac{(r\sin\varphi-z_0)^2}{2\sigma_3^2}\right)} \cos\varphi d\varphi \qquad (60).$$

The integration can be carried out numerically, but it can be performed very easily in closed form for $x_0=y_0=z_0=0$. In that case we have, assuming for simplicity $\sigma_1=\sigma_2=\sigma_3=\sigma$

$$p(r) = \frac{r^2}{\sigma^3\sqrt{(2\pi)^3}} \int_0^{2\pi} d\theta \int_{-\frac{\pi}{2}}^{\frac{\pi}{2}} e^{-\frac{(r\cos\varphi\cos\theta)^2+(r\cos\varphi\sin\theta)^2+(r\sin\varphi)^2}{2\sigma^2}} \cos\varphi d\varphi \qquad (61)$$

or

$$p(r) = \frac{r^2}{\sigma^3\sqrt{(2\pi)^3}} \int_0^{2\pi} d\theta \int_{-\frac{\pi}{2}}^{\frac{\pi}{2}} e^{-\frac{r^2}{2\sigma^2}} \cos\varphi d\varphi \qquad (62)$$

$$p(r) = \frac{2r^2 e^{-\frac{r^2}{2\sigma^2}}}{\sigma^3\sqrt{(2\pi)^3}} \int_0^{2\pi} d\theta = \frac{2r^2 e^{-\frac{r^2}{2\sigma^2}}}{\sigma^3\sqrt{2\pi}} \qquad (63).$$

Eq. (63) is the extension to the 3D configuration of the well known Rayleigh distribution, valid for the 2D case [19].

In Fig. 18 the expected radial distributions of point like events located at $r$=0, $r$=0.4 and $r$=0.8 in a generic scintillator vessel of 1m radius are reported.

The sigma for the $x$, $y$ and $z$ variables assumed in the calculation is equal to 0.15 m. For events located in the center the most probable estimated radial position is .0.21 m, for events located at 0.4 m is 0.45 m, and for events located at 0.8 is 0.83. So, the more the events are far from the centre, the more the peak position of the radial distribution is close to the true event location.

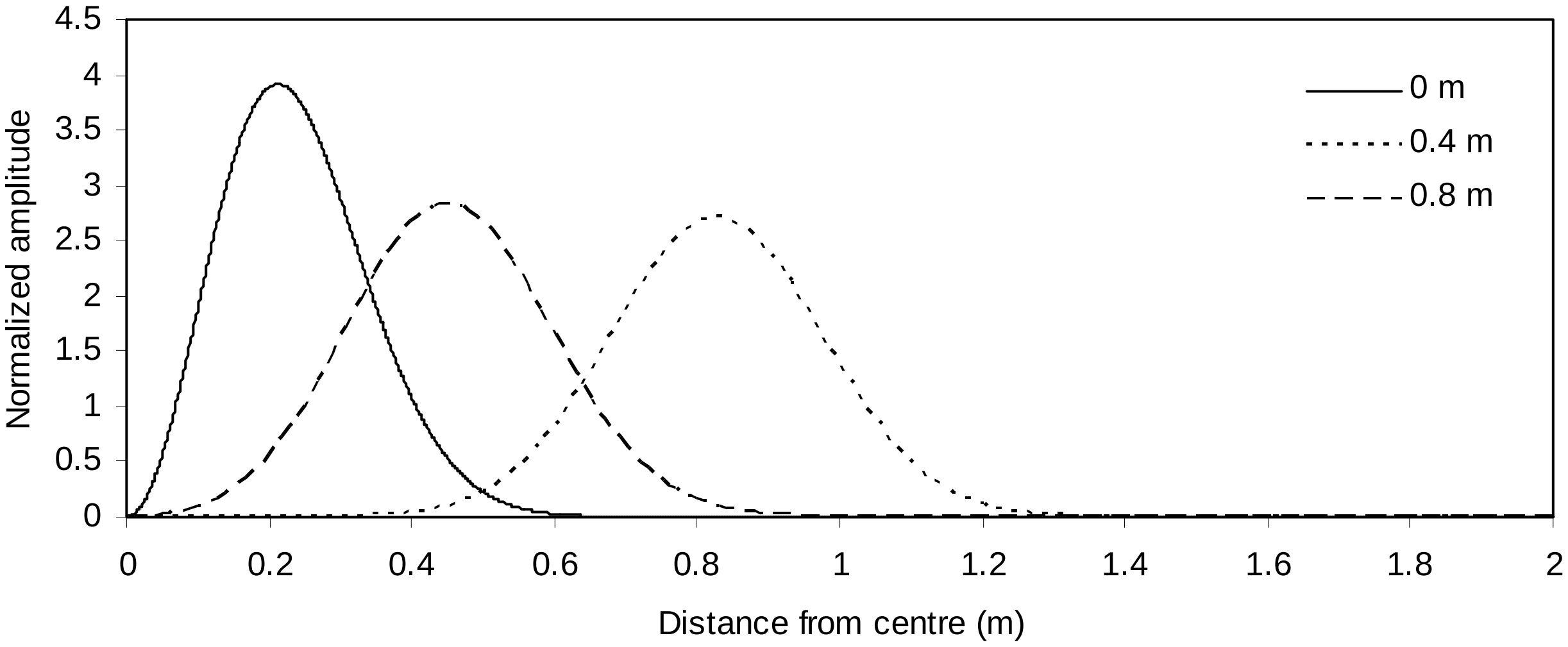

*Fig. 18 – Estimated radial distributions of point-like events at three different true locations from the centre of a detector of radius 1 m and for $\sigma$=0.15 m*

Eq. (63) is applicable also in the case in which it is needed to perform spatial correlation of different events. In particular, if two events originate from the same location (as in special radioactive decay correlated sequences) the radial distance of the corresponding estimates obeys to relation (63), with the caveat that the sigma to be introduced is that affecting the individual coordinates multiplied by $\sqrt{2}$.

## 11. Reconstructed distribution of uniform events

From relation (60) it can be inferred the radial distribution of events uniformly distributed in the vessel. For this purpose it is enough to multiply eq. (60) by the density of the events as function of $r_0$ (the true radial event position, scanning the whole vessel) and then integrate the result over $r_0$ itself. Furthermore, because of the spherical symmetry, it can be assumed without loss of generality $x_0$=$r_0$, $y_0$=0 and $z_0$=0. Hence, since the probability density function of $r_0$ is

$$p(r_0) = \frac{3}{r_v^3} r_0^2 \qquad (64)$$

we have

$$p(r) = \int_0^{r_v} dr_0\, p(r_0) \frac{r^2}{\sigma_1\sigma_2\sigma_3\sqrt{(2\pi)^3}} \int_0^{2\pi} d\theta \int_{-\frac{\pi}{2}}^{\frac{\pi}{2}} e^{-\left(\frac{(r\cos\varphi\cos\theta - r_0)^2}{2\sigma_1^2} + \frac{(r\cos\varphi\sin\theta)^2}{2\sigma_2^2} + \frac{(r\sin\varphi)^2}{2\sigma_3^2}\right)} \cos\varphi d\varphi \qquad (65).$$

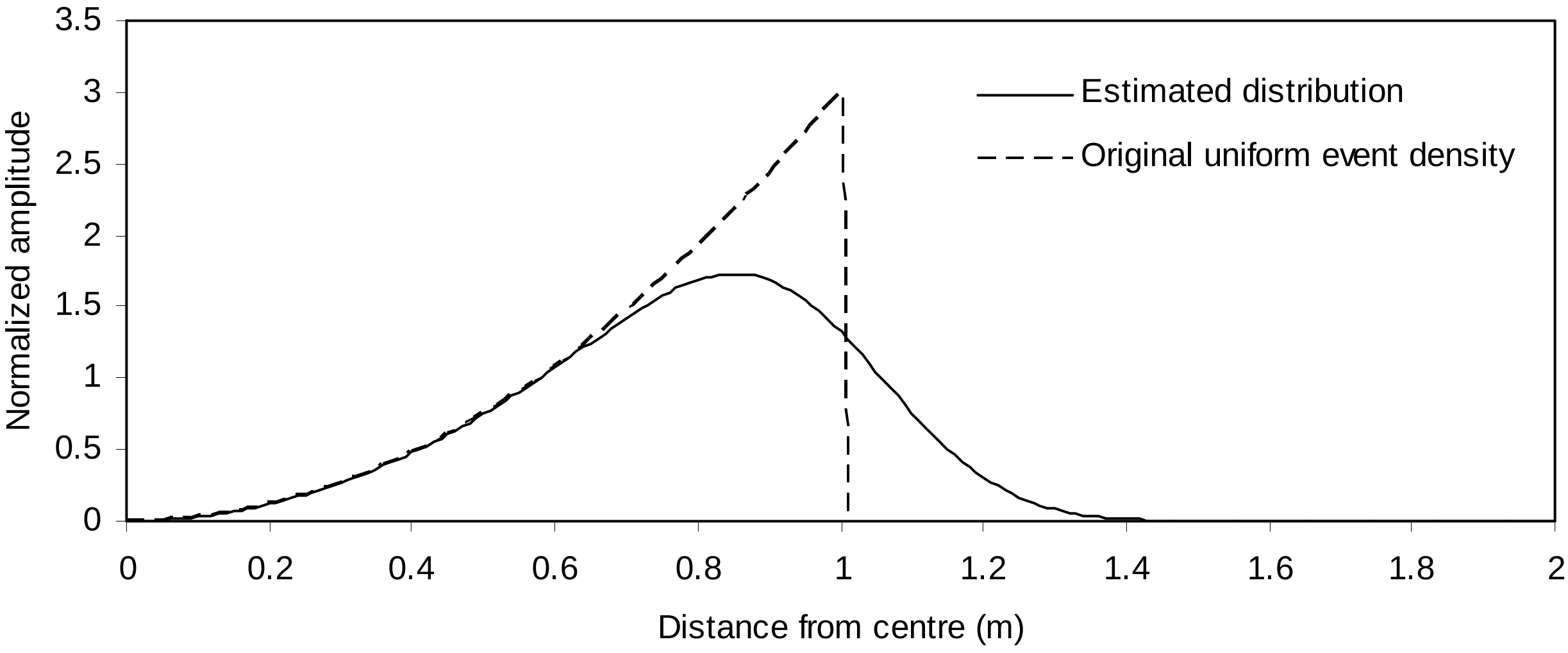

*Fig. 19 – $r^2$ uniform event density and the corresponding estimated radial distribution evaluated for a detector of 1 m radius and a resolution of 0.15 m*

The expected distribution stemming from eq. (65) is reported in Fig. 19 for the case of the CTF vessel.

In Fig. 19 the $r^2$ geometrical distribution is also displayed. Interestingly, in the core of the vessel the two distributions coincide, while it is close to the boundary that the effect of the finite resolution manifests with a tail of events whose location is estimated outside the physical region of the containment vessel.

## 12. Reconstructed distribution of external events

A calculation similar to that of previous paragraph can lead to the evaluation of the estimated distribution of external events (external in the sense that they are induced inside the scintillator by gamma rays originated by radioactive contaminants in the surrounding medium), in the reasonable assumption that the actual position of such events follow an exponential decaying profile inward from the vessel wall. Thus in this case the probability density function of these events can be written as

$$p(r_0) = \frac{r_0^2 e^{-\frac{r_v - r_0}{\lambda}}}{\int_0^{r_v} r_0^2 e^{-\frac{r_v - r_0}{\lambda}}} \qquad (66)$$

and the required distribution is obtained from eq (65) in which it is inserted the relation (66) as $p(r_0)$.

The expected distribution is reported in Fig. 20 for the case of the CTF vessel; for comparison it is reported again the expected distribution for uniformly distributed internal events.

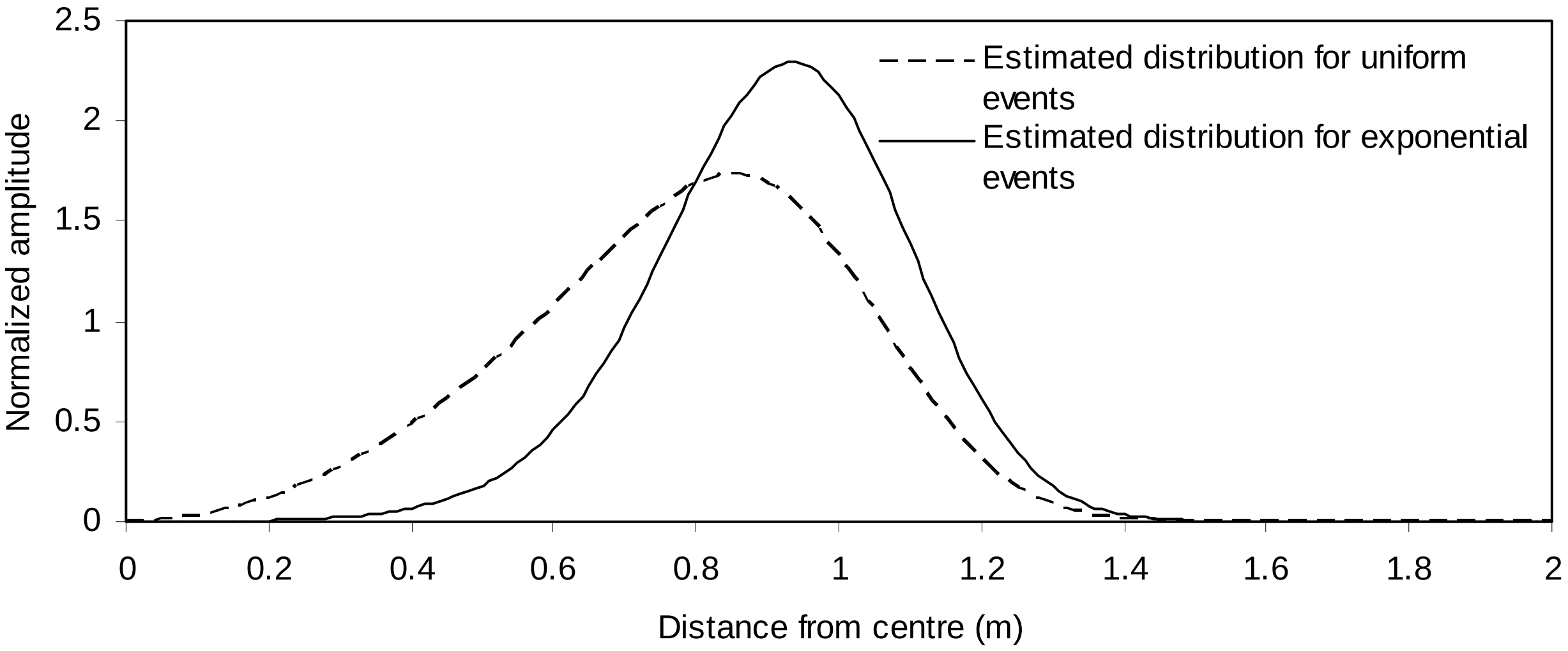

*Fig. 20 – Estimated event densities for uniformly (internal) and exponentially (external) distributed events evaluated for a detector of 1 m radius and a resolution of 0.15 m*

**13. Conclusions**

The time and the associated event position distributions in scintillation detectors can be accurately studied through detailed Monte Carlo modeling. In the special case of spherical geometry, however, useful formulas can be derived to give an approximate analytical description of the detector signal responses, which can usefully complement the more precise Monte Carlo outputs. In this paper we have given a thorough account of their derivation for several cases of practical interest.

**Acknowledgement**

The authors acknowledge the support of INFN to this research.